%% file: main.tex
\documentclass{article}

\usepackage[english]{babel}

 \usepackage[a4paper,top=2cm,bottom=2cm,left=3cm,right=3cm,marginparwidth=1.75cm]{geometry}

\usepackage{amsmath}
\usepackage{bm}
\usepackage{authblk}
\usepackage{graphicx}
\usepackage[colorlinks=true, allcolors=blue]{hyperref}
\usepackage{soul}
\usepackage{xcolor}
\usepackage{tabularx}
\usepackage{threeparttable}	
\usepackage{threeparttablex}
\usepackage{array}
\usepackage{makecell}
\usepackage{siunitx}
\usepackage{subfigure}
\usepackage{subcaption}
\usepackage{multirow}

\newcommand{\marka}{\color{black}}

\title{Initial performance results of the JUNO detector}
\input{author}

\begin{document}

\maketitle

\begin{abstract}
The Jiangmen Underground Neutrino Observatory (JUNO) started physics data taking on 26~August~2025. JUNO consists of a 20-kton liquid scintillator central detector, surrounded by a 35~kton water pool serving as a Cherenkov veto, and almost 1000~m$^2$ of plastic scintillator veto on top. The detector is located in a shallow underground laboratory with an overburden of 1800~m.w.e. This paper presents the performance results of the detector, extensively studied during the commissioning of the water phase, the subsequent liquid scintillator filling phase, and the first physics runs. The liquid scintillator achieved an attenuation length of 20.6~m at 430~nm, while the high coverage PMT system and scintillator together yielded about 1785 photoelectrons per MeV of energy deposit at the detector centre, measured using the 2.223~MeV $\gamma$ from neutron captures on hydrogen with an Am-C calibration source.
The reconstructed energy resolution is 3.4\% for two 0.511~MeV $\gamma$ at the detector centre and 2.9\% for the 0.93~MeV quenched $^{214}$Po alpha decays from natural radioactive sources. The energy non-linearity is calibrated to better than 1\%. Intrinsic contaminations of $^{238}$U and $^{232}$Th in the liquid scintillator are below 10$^{-16}$~g/g, assuming secular equilibrium. The water Cherenkov detector achieves a muon detection efficiency better than 99.9\% for muons traversing the liquid scintillator volume.
During the initial science runs, the data acquisition duty cycle exceeded 97.8\%, demonstrating the excellent stability and readiness of JUNO for high-precision neutrino physics.

\end{abstract}


\section{Introduction}
The Jiangmen Underground Neutrino Observatory (JUNO) was proposed to determine the Neutrino Mass Ordering~(NMO)~\cite{Zhan:2008id,Zhan:2009rs}.
The detector is located at a shallow depth in an underground laboratory under the Dashi Hill (about 650~m of overlying rocks, corresponding to 1800~m.w.e.), very close to Jinji town, 43~km south-west of Kaiping city, in the prefecture-level city Jiangmen, Guangdong province, China.
The location has been chosen to be at an equal 52.5~km distance from the Yangjian and Taishan nuclear power plants, optimized for measuring the NMO~\cite{JUNO:2015zny,JUNO:NMOsensitivity:2025}.
JUNO consists of a 20-kton Liquid Scintillator~(LS) Central Detector~(CD), a 35~kton Water Cherenkov veto Detector~(WCD), and a 1000~m$^2$ plastic scintillator Top Tracker~(TT).
With an unprecedented large target mass and high energy resolution, JUNO will usher in a new era of precision measurements in the neutrino sector, aiming world-leading sub-percent precision on the oscillation parameters $\Delta m_{21}^2$, $\Delta m_{31}^2$, and $\sin^2\theta_{12}$~\cite{JUNO:NuOscPar:2022}.
As discussed in these publications~\cite{JUNO:2015zny,JUNO:DetPhys:2022}, JUNO has a very rich physics program, since it will be possible to detect neutrinos and anti-neutrinos produced from terrestrial and extra-terrestrial sources:
supernova burst neutrinos~\cite{JUNO:CCSN:2024} and diffuse supernova neutrino background~\cite{JUNO:DSNB:2022}, geo-neutrinos~\cite{JUNO:geoneutrino:2025}, atmospheric neutrinos~\cite{JUNO:nuatmLowEnergy:2021}, and solar neutrinos~\cite{JUNO:solar8B:2021,JUNO:solar:2023,JUNO:2022jkf}.
Finally, JUNO is sensitive to physics searches beyond the Standard Model of Particle Physics, such as studies of proton
decay in the $p \rightarrow K^+ \overline{\nu}$ channel~\cite{JUNO:pdecay:2023}, neutrinos generated by dark matter annihilation in the Sun or in the galactic Halo~\cite{JUNO:DarkMatterGalacticHalo:2023}, and other non standard interactions.
An extended overview of the JUNO physics goals can be found elsewhere~\cite{JUNO:DetPhys:2022}.

The civil construction of the JUNO underground lab started in January 2015.
The installation of the Stainless Steel support structure started on 19~December~2021 while the first panel of the acrylic vessel was installed on 27~June~2022.
The installation of the photomultiplier tube (PMT) modules and electronics boxes started on 23~October~2022 and finished on 17~December~2024.
After a thorough cleaning of the inner surface of the acrylic vessel, CD and WCD were filled with purified water between 18~December~2024 and 2~February~2025.
The water in the CD was exchanged with purified LS between 8~February~2025 and 22~August~2025.
The detector commissioning was carried out simultaneously with the water filling and LS exchange.
The TT started to be assembled on 7~January~2025, after the top of the WCD was closed, and it was completed on 11~June~2025.
After several days of detector calibration, the physics data taking started on 26~August~2025.
This paper will briefly introduce the primary detector components as well as the filling and commissioning efforts, followed by the first detector performance results.

\section{The JUNO detector}
\begin{figure}[htbp]
\centering
\includegraphics[width=0.75\columnwidth]{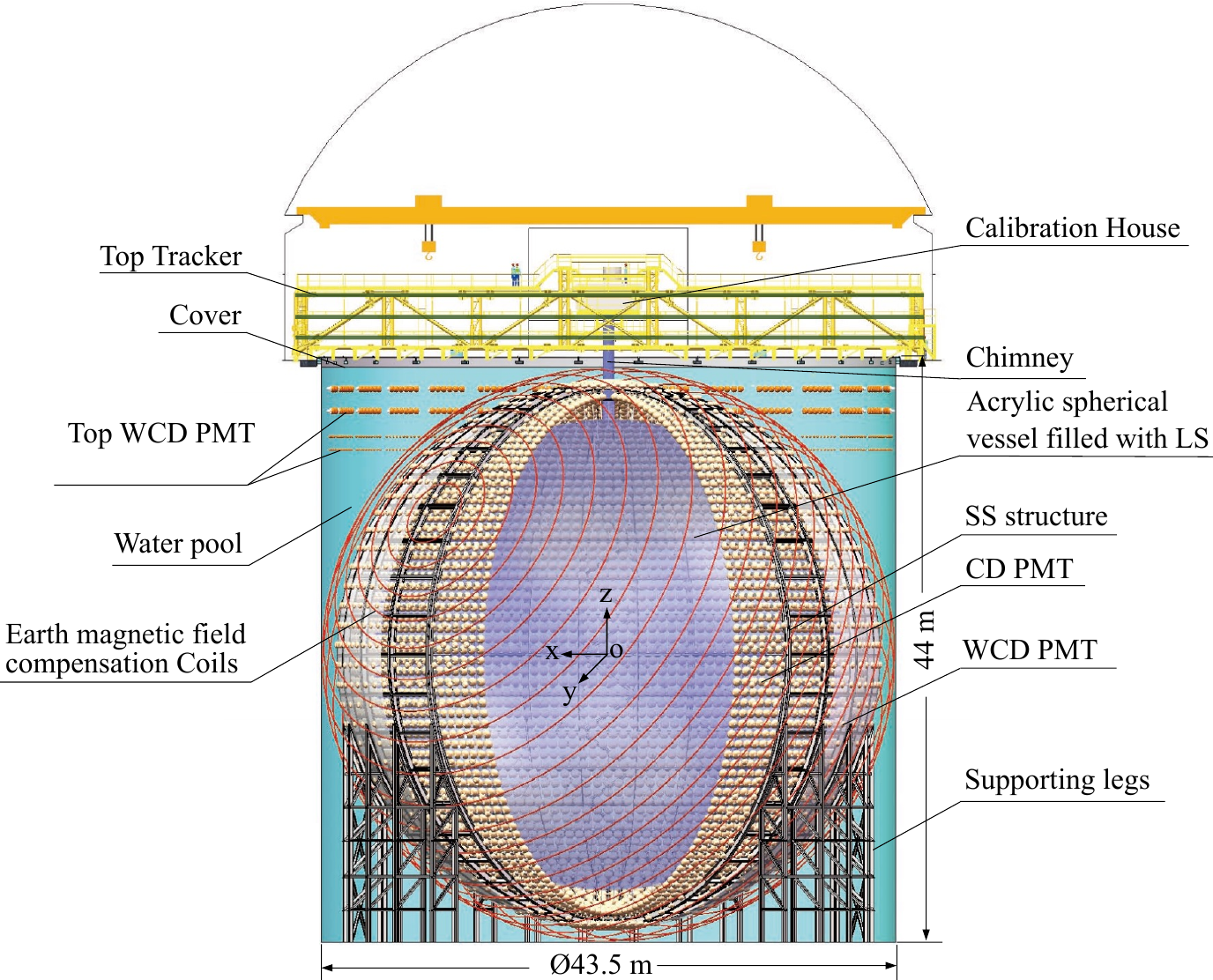}
  \caption{\label{fig:JUNO:CAD}Schematic view of the JUNO detector. The spherical Central Detector (CD) contains 20 ktons of organic Liquid Scintillator (LS) enclosed in an acrylic vessel, serving as neutrino target. It is surrounded by a cylindrical Water Cherenkov Detector (WCD) filled with ultra-pure water. The Top Tracker (TT) positioned above the setup measures crossing muon tracks with high precision.}
\end{figure}

A schematic drawing of the JUNO detector can be seen in Figure~\ref{fig:JUNO:CAD}: where the CD, the WCD, and the TT are explicitly indicated. Also shown is the CD chimney, which connects the CD with the calibration house for the calibration of the detector.
The JUNO reference system is show in Figure~\ref{fig:JUNO:CAD} with the origin of the axes in the CD centre.
In the following sections only the components relevant to the presented results will be discussed in details. Further
details on all JUNO hardware and physics potentials can be found here~\cite{JUNO:DetPhys:2022}.

\subsection{The JUNO Central Detector}
The CD is a 20~kton LS target contained in a spherical
acrylic vessel with an inner diameter of 35.4~m. The vessel has been built by bonding, in place,  263 pieces of highly transparent spherical panels, with a 120~mm thickness; it is kept in position by means of 590
bars that connect the acrylic sphere to a stainless steel (SS) structure which supports the CD. The SS structure has a spherical shape with an inner diameter of 40.1~m
and serves multiple purposes: to safely support the CD
during the construction, filling and operational phases, and to support the JUNO photon detector systems and their readout electronics.

Photon detection in JUNO is realized with two independent systems, with the photocathode directed towards the centre of the CD.
The main system consists of 17596 20-inch PMTs, (referred in the following as `large PMTs') installed on the SS structure.
Two types of PMTs have been employed:
4939 dynode-PMTs (model R12860-50, produced by Hamamatsu Photonics K.K. (HPK), Japan)
and 12657 Micro-Channel Plate photomultipliers, MCP-PMTs, (model GDB6201, produced by Northern Night Vision Technology Co. (NNVT), China).
Details on the testing and acceptance procedure of the 20-inch PMTs can be found here~\cite{JUNO:PMTmassTesting:2022}.
In addition, 25587 3-inch PMTs (referred in the following as `small PMTs') have been produced from Hainan Zhanchuang Photonics Technology (China) and installed in the space between the 20-inch PMTs.
The small PMTs constitute a complementary photodetection system with a wider dynamic range than the large PMTs. They enhance JUNO's overall performance by improving calibration precision and mitigating instrumental non-linearities~\cite{JUNO:CalibStrategy:2021}.
Details on the testing and acceptance program of the 3-inch PMTs can be
found here~\cite{Cao_2021,JUNO:sPMTinstrumentation:2025}.

\subsection{The JUNO Veto System}
The CD is complemented with two independent veto detectors designed to provide an efficient discrimination and reduction against environmental radioactivity and background related to
residual cosmic muons crossing the LS target.

The water pool is a cylinder with 44~m height and 43.5~m diameter and
consists of the WCD with 35~kton active mass and remaining 5~kton of water located between the outer surface of the acrylic vessel and the SS structure supporting the PMTs, and forming a buffer region for the photomultiplier system.
The walls and ground floor of the cylinder are covered by a 5~mm thick High Density Polyethylene liner to reduce radon diffusion from the rock walls to the water pool.

The Cherenkov photons produced by charged particles travelling through water are detected
by 2404 20-inch MCP-PMTs positioned on the SS structure and facing the outer walls of the water pool. The WCD is optically separated from the CD thanks to
Tyvek\textsuperscript{\textregistered} reflective foils covering
the complete outer lateral surface of the SS sphere.
In addition, Tyvek\textsuperscript{\textregistered} foils completely cover
the water pool walls and basement to increase the light collection efficiency.
To further improve the Cherenkov light collection acceptance, four additional rings of PMTs have been installed on the water pool walls.
Two rings at $z = 20.66~\mbox{m}$ and $z = 18.98~\mbox{m}$
realized with 348 20-inch MCP-PMTs and two rings at $z = 17.05~\mbox{m}$ and $z = 16.05~\mbox{m}$ made with 600 8-inch PMTs (from the former Daya Bay experiment~\cite{DayaBay:detector-paper:2015}), with respect to the CD centre. A system of 135 LED flashers which are deployed on the wall and the bottom of the water pool provides a timing calibration of the PMTs.

The top of the water pool is covered by a black rubber layer and sealed using gas-tight zippers. This covering layer is placed about one meter higher than the WCD water level; the residual volume is filled with nitrogen with a slight overpressure (few hundreds Pa) with respect to the experimental hall atmosphere to prevent radon contamination entering the water pool.

Coils are mounted on the SS structure to compensate for the Earth's magnetic field and minimize its impact on the photoelectron collection efficiency of the PMTs. The average residual Earth's magnetic field strength within the volume surrounded by the coils is about 0.05~G.
Impact on the PMT detection efficiency of PMTs is smaller than 1\% according to the measurement published in~\cite{Zhang:2021ikn}.

On top of the water pool, an array of plastic scintillator detector, the TT~\cite{JUNO:TTnim:2023}, has been installed.
It covers about 60\% of the top surface of the WCD and it has been designed
to precisely measure muons crossing the detector.
These tracks can then be used as a reliable muon sample to independently tune and validate the track reconstruction in the CD and WCD.
Moreover these tracks can also be used to refine the constraint on cosmogenic backgrounds as well as fast neutrons induced by muons traversing the rock surrounding the WCD by looking for a correlation in space and time with the CD and
WCD.
As described in detail in~\cite{JUNO:TTnim:2023}, the TT is composed of 63 `walls', distributed in 3 vertical layers of $7 \times 3$ walls placed horizontally in a grid pattern.
Each wall is constructed by placing horizontally in a support structure
8 modules, each with 64 plastic scintillator strips, in 2 levels aligned in perpendicular directions.
The light produced by a muon crossing the plastic scintillator is collected by a
wavelength shifting optical fiber and directed to a specific channel of each
multi-anode PMT (model H7546, produced by HPK, Japan) located at each end of the fiber. Each wall has a sensitive surface of about $6.8~\mbox{m}\times 6.8~\mbox{m}$ with a granularity to identify the
crossing position of a muon of $2.64~\mbox{cm}\times 2.64~\mbox{cm}$.
Thanks to this granularity and the distance of consecutive layers of TT walls
(about 1.5 m in most of the surface), the TT is expected to achieve a median angular
resolution of $0.2^\circ$ to track muons.
While the TT modules had been fabricated for the OPERA experiment~\cite{OPERA:TT},
the electronics used to read out all 63488 channels and supporting structure of the TT were redesigned for JUNO to account for the different operation conditions.

\subsection{Auxiliary JUNO facilities}
Several auxiliary plants and facilities have been constructed to support the preparation, installation, and operation of the JUNO detector. The following sections provide a brief overview of those systems relevant to the detector filling and commissioning phases.

\subsubsection{The water plants}
The water supply for the WCD and the other major experimental needs is provided by a dedicated plant designed to provide High-Purity Water (HPW) with high fluxes, up to 100~tons/hour. An additional Ultra Pure Water (UPW) plant~\cite{JUNO:upwater:2024}, with smaller capacity (4~tons/hour), has been designed to reduce radioactive contaminants, such as $^{222}\mbox{Rn}$ that could contribute to the detector background~\cite{JUNO:2021kxb}.
The UPW plant employs an online radon removal system, that thanks to micro-bubble generators and multistage degassing membranes~\cite{JUNO:upwater:2024}, shows a greater than 99.9\% $^{222}$Rn removal efficiency, reducing the radon concentration in water down to $1~\mbox{mBq}/\mbox{m}^3$ satisfying the stringent JUNO design requirements~\cite{JUNO:DetPhys:2022} for the Water Extraction plant and for the OSIRIS detector, which are both described in the following sections.
The HPW system has been operated continuously during the detector filling phase and will be operated in recirculation mode during detector running phase. In addition, a very sensitive online radon concentration monitor~\cite{JUNO:upwater:2024}, capable of detecting very low radon concentrations is in operation to monitor the radon concentration in water.

In this water system, online measurements of the resistivity and oxygen concentration, and sampling measurements of particulate matter and radium concentration of the ultrapure water, were performed. The resistivity remained consistently above 18.16 M$\Omega\cdot\mbox{cm}$, the dissolved oxygen concentration ranged between 1 and 2 ppb, the particulate count was approximately 100 particles/L, and the radium concentration was measured and maintained below $4~\mu\mbox{Bq}/\mbox{m}^3$ and measured with a self-developed apparatus~\cite{Li:2024lcy}.

\subsubsection{The LS purification system}
\label{LS_Purification}
The 20,000 tons of LS to be filled into the Central Detector (CD) constitute the active target of the JUNO experiment.
The selected LS mixture consists of four components, in different concentrations:
the core is made of linear alkyl benzene (LAB) as solvent, doped with 2,5-diphenyloxazole (PPO) as primary fluor,
1,4-bis(2-methylstyryl)benzene (bis-MSB) as wavelength shifter, and butylated hydroxytoluene (BHT) as antioxidant. After several tests done at the Daya Bay laboratory, the final concentration of the fluors was determined to be 2.5 g/L PPO and 3 mg/L bis-MSB, to be diluted in purified LAB~\cite{JUNO-DYB:LScomposition:2021}. It was successively decided to dose also about 42.7 mg/L BHT, in order to prevent long-term optical degradation and ensure an excellent transparency of the LS during the expected 20-year JUNO lifetime.

Even if the LAB and the powders have been supplied by specialized companies with low levels of U and Th contaminations, a complete system of purification plants has been designed and constructed to improve the optical and radiopurity properties of the LS, right before filling the CD.
Starting from the raw materials delivered by the suppliers, the following purification steps are carried out at the JUNO site to produce the final LS mixture:
\begin{itemize}
\item the raw LAB is firstly purified above ground by filtering it through alumina powder ($\mbox{Al}_2\mbox{O}_3$), in order to enhance light transmittance and improve transparency~\cite{Zhu:OpticalPurifiction:2022};
\item the purified LAB is sent to a distillation plant operated under partial vacuum~\cite{Landini:distillation:2024}, to discard high boiling contaminants such as U, Th and K compounds;
\item the third step performs mixing of the purified LAB solvent with PPO, bis-MSB and BHT solutes in higher concentration, producing the so-called
  LS master solution. An additional water washing of this solution is performed to purify PPO and bis-MSB mainly from U and Th.
  After purification, the master solution is further diluted with distilled LAB until the final JUNO recipe is obtained and then transported to the underground laboratory by means of a dedicated 1.3~km long stainless-steel pipe.
\end{itemize}

Two additional purification plants operate in the underground laboratory in a dedicated
hall, right before delivering the purified scintillator to the CD:
\begin{itemize}
\item the water extraction plant~\cite{Ye:WaterExtraction:2021} is used to further reduce the amount of polar contaminants and other residues containing U/Th compounds, especially those introduced by PPO and bis-MSB;
\item gas stripping~\cite{Landini:distillation:2024} with high purity nitrogen~\cite{Ling:NitrogenPlant:2024} is the final stage of the purification procedure and is very effective in removing radioactive gases ($^{222}\mbox{Rn}$, $^{85}\mbox{Kr}$, $^{39}\mbox{Ar}$) and remaining gaseous impurities (for instance oxygen, which could cause quenching in the LS).
\end{itemize}

All these plants have been tested and optimized during several joint commissioning campaigns from 2023 until the start of the LS filling, together with OSIRIS for monitoring early radiopurity levels.

\subsubsection{The OSIRIS system}

As a final stage of LS radiopurity control, the pre-detector OSIRIS (Online Scintillator Internal Radioactivity Investigation System) can monitor ton-scale samples of LS for radon, $^{238}$U, $^{232}$Th, $^{210}$Po and $^{14}$C levels \cite{JUNO:2021wzm}. The system is based on a 20-ton scintillator volume, surrounded by 80 large PMTs and shielded by a 9-by-9-meter cylindrical water tank. Background levels are assessed by measuring decays based on their scintillation signals: Bi-Po coincidence analyses, both for $^{214}\mbox{Bi}$-$^{214}\mbox{Po}$ and $^{212}\mbox{Bi}$-$^{212}\mbox{Po}$ cascade decays, allow to extract Bi-Po rates and time development, while spectral analysis is used for $^{210}\mbox{Po}$ and $^{14}\mbox{C}$. Given LS batch size and background levels, the maximum sensitivity that OSIRIS could achieve in extended runs was better than $10^{-15}$\,g/g for $^{238}$U and $10^{-16}$\,g/g for $^{232}$Th \cite{osiris:ana:2025}.

\subsubsection{The filling, overflow and circulation system}
The Filling, Overflow and Circulation (FOC) system is an auxiliary equipment of the JUNO CD which is composed of one storage tank and two overflow tanks, each with a volume of $50~\mbox{m}^3$, an ultrapure nitrogen flushing system and several tanks, pipelines, circulation pumps and monitoring sensors to manage both purified water and liquid scintillator inside the detector.
The tanks put in connection the purification plants producing the LS with the detector acrylic sphere through the top chimney. Moreover, the FOC delivers high purity water to the CD and pure water from the HPW plant to the WCD.

It has been designed with three main functions:
\begin{enumerate}
\item synchronous filling of the CD and WCD with high-purity water during the first JUNO filling phase, and replacing water in the CD with LS during the second JUNO filling phase;
\item circulate the LS from the detector passing through the underground LS purification system, in case online re-purification is needed;
\item control and stabilize the LS level in the CD within 20~cm following the LS temperature changes in the range
$(21 \pm 1.4)^{\circ}\mbox{C}$ during the JUNO running phase, by means of the overflow tanks.

\end{enumerate}

\subsection{The calibration system}
Energy and timing calibration are key ingredients for the proper understanding and analysis of JUNO data.
Moreover, according to the design goals and expected sensitivities~\cite{JUNO:NMOsensitivity:2025}, the energy scale of the LS must be known to better than one percent.
As described here~\cite{JUNO:CalibStrategy:2021}, JUNO has developed a complex and multiform strategy to determine the energy scale and correct non-linearities and non-uniformities in the detector response.
It consists of several independent pieces of calibration hardware that are able to place different sources in different positions along the central axis of the CD, on a circle at the LS-acrylic vessel boundary, and in the region in between have been designed and built. These are:
\begin{itemize}
\item the Automatic Calibration Unit (ACU)~\cite{calib:ACU},
      which allows to deploy multiple radioactive sources, a laser source, or auxiliary sensors, such as
      a temperature sensor, one at a time, along the CD central axis;
\item the Guide Tube (GT) calibration system~\cite{calib:GT} has been       designed to deploy a radioactive source along a given longitude on the outer surface of the acrylic vessel;
\item the Cable Loop System (CLS)~\cite{calib:CLS} allows to access off-axis calibration positions to investigate
      non-uniformities of the detector response. Adjusting the length of two connecting cables, the system can
      move a source on a vertical half-plane, covering about 79\% of a CD vertical plane with a positional repeatability better than 10~mm~\cite{calib:CLS}. An ultrasonic positioning system is deployed in the CD to determine the source location with a 3~cm precision~\cite{JUNO:USS1,JUNO:USS2};
\item the Remotely Operated Vehicle (ROV)~\cite{JUNO:ROV} allows to
      deploy radioactive sources throughout the full detector volume. It is a compact cylindrical module equipped with jet pumps, a vertical positioning device, and an umbilical cable that provides power, control, and mechanical support. Guided by combined ultrasonic and pressure-sensor feedback, the ROV can be moved to nearly any location inside the central detector, offering broad spatial coverage and useful cross-checks with other calibration subsystems.
\end{itemize}

\section{JUNO filling and LS properties}
Filling and commissioning of the JUNO detector were carried out simultaneously from 18~December~2024 to 22~August~2025.
To monitor the in-situ LS quality and avoid Radon leakage to LS, the JUNO detector has kept running since 3~February~2025.
Large PMT waveforms in the global trigger window are readout and reconstructed by the data acquisition system~(DAQ), followed by the online event reconstruction for vertex and energy determination.
Cascade decays from $^{214}$Bi and $^{214}$Po are selected to provide a real-time monitoring of $^{222}$Rn levels in the CD.
Thanks to the careful leakage checks and the nitrogen protection of all pumps and valves of the LS purification and filling systems, and the real-time $^{222}$Rn level monitoring, the average $^{222}$Rn contamination in the fresh LS is smaller than 1~mBq/m$^3$, much better than the required 5~mBq/m$^3$.

\subsection{JUNO detector filling}

Prior to filling,  the CD acrylic vessel was cleaned in two steps. First of all a large amount of water mist was generated inside the CD through a self-developed water mist generation device, improving the air cleanliness inside the CD  from class 10000 to class 100. Then, the inner surface of the acrylic vessel is rinsed thoroughly with water sprayed by a custom 3D rotating nozzle capable of flushing water with high pressure. The second step cleaning judgement standard is to test and compare the water quality of the inlet water and outlet water, with monitoring items including water particle size, water absorbance, and ICP-MS water quality detection. The water used in both steps is produced by the ultra pure water system.

The whole filling process was carefully performed by the FOC system, which is responsible for filling both the CD and the WCD while controlling liquid level synchronization and pressure balance, to ensure the structural integrity of the CD.

The filling procedure has been structured in two main phases:
\begin{itemize}
    \item a simultaneous filling of the CD and the WCD with pure water is first performed, to balance buoyancy and mechanical stresses on the CD acrylic vessel;
    \item afterwards, the exchange between water and scintillator inside the CD is gradually and smoothly exploited by pumping water out from the bottom of the CD and loading newly produced scintillator from the top.
\end{itemize}

\subsubsection{Pure water filling phase}
The water filling process for the JUNO CD is a meticulously controlled operation designed to ensure structural integrity and safety, while preparing for the subsequent scintillator filling. The process involved synchronous filling of both the CD and WCD, with stringent requirements for water purity and liquid levels management, to avoid pressure imbalances that could exceed the safety thresholds for the CD acrylic vessel.

The water produced by the HPW plant met exceptionally high purity standards, including U and Th content at or below
$10^{-15}~\mbox{g}/\mbox{g}$, $^{222}$Rn concentration below $10~\mbox{mBq}/\mbox{m}^3$ and $^{226}$Ra concentration
under $50~\mu\mbox{Bq}/\mbox{m}^3$~\cite{JUNO:upwater:2024}.
Additional specifications require water resistivity exceeding $18~\mbox{M}\Omega\cdot\mbox{cm}$, oxygen content below 10~ppb, and particle levels compliant with JUNO's rigorous cleanliness standards.
To guarantee these parameters, the water undergoes supplementary purification via a super filter before entering the CD.

The filling process incorporated comprehensive structural monitoring with displacement sensors tracking CD components (chimneys, TT bridge, and stainless-steel grid structure) and 20\% of the acrylic vessel's support rods equipped with load cells and temperature sensors, all monitored in real-time by operators.

Safety and monitoring systems were the backbone of this complex operation. Five redundant level meters provided precise measurements, with CD monitoring achieving $\sim$0.2\% accuracy and WCD measurements maintaining $\sim$0.2\% precision. The entire process followed strict level difference limits derived from finite element analysis, with automatic activation of an 'on-off' safety mode should any parameters exceed predetermined alarm thresholds.
These comprehensive measures ensured both operational safety and structural
integrity throughout the filling process.

The successful completion of this massive undertaking saw approximately 64,000 tons of pure water filling the whole system to a height of 43.5~m within just 46 days, from 18~December~2024 until 2~February~2025.
While a minor calibration issue with the WCD level gauge was noted, continuous rod force monitoring confirmed the detector's structural stability throughout the operation.
This achievement not only prepared the CD for subsequent LS filling but also established the necessary ultra-low contamination environment crucial for JUNO's scientific mission.

\subsubsection{Liquid scintillator filling phase}
The principal and most time-consuming phase of the detector filling procedure
concerned the replacement of pure water by the LS produced and purified online at the JUNO facility. As during the water phase, the structural integrity of the CD must be preserved, also taking into account for the different densities of the two liquids
($0.856~\mbox{g}/\mbox{cm}^3$ for LS and roughly $1~\mbox{g}/\mbox{cm}^3$ for water),
which leads to increasing buoyancy imposed insisting on the vessel and the connecting bars as the exchange progresses. To compensate for the density gap and keep the pressure difference within the allowed ranges, the liquid level inside the acrylic vessel was progressively raised during the LS exchange process. Several parameters, including level and pressure balances, position or stress variations of the connecting rods, temperatures and flow-rate, were continuously monitored to ensure a smooth filling and detector safety.
Water was removed using a self-priming pump connected at the bottom chimney. Simultaneously, the purified liquid scintillator was loaded from the top chimney at $7~\mbox{m}^3/\mbox{h}$, resulting in about $168~\mbox{m}^3/\mbox{day}$ and 6 months scheduled to complete this phase.

The highly demanding quality required in terms of radiopurity and optical features of the JUNO liquid scintillator had to be preserved also passing the FOC system and inside the vessel for the whole lifetime of JUNO.
Specific treatments and precision cleaning of all internal surfaces, together with a careful material selection and screening~\cite{JUNO:RadContrStrategy:2021},
were adopted to avoid contamination of the LS.


To avoid $^{222}$Rn leakage into LAB or LS during the six-months LS production and filling period, several measures were taken into account:
\begin{itemize}
\item During production and transportation, the LAB was sealed under nitrogen gas at 0.8 bar in custom transport tanks. The on-site storage tank was likewise maintained under nitrogen protection. The PPO and bis-MSB were enclosed in double-layer, vacuum-sealed bags to ensure containment integrity, such that failure of one layer would not compromise the other.
\item All components of the liquid scintillator purification and filling systems were required to have a leak rate below 10$^{-8}$ mbar$\cdot$L/s. Flanges were sealed with double O-rings, and the intermediate space was purged with nitrogen for additional protection. Certain valves and pumps were further safeguarded by external nitrogen-filled enclosures.
\item The acrylic sphere's upper chimney and the calibration housing are continuously purged with nitrogen to maintain a slight positive pressure, preventing the infiltration of radon-rich external air into the acrylic vessel.
\item During the LS filling process, a continuous high-purity nitrogen flow ($\sim10~\mbox{m}^3/\mbox{h}$) was maintained in both the tanks and calibration house to establish positive pressure barriers against radon intrusion, with dedicated sensors monitoring the positive pressure.
\item Real-time $^{222}$Rn activity is monitored by tagging $^{214}$Bi-$^{214}$Po cascade decays during data taking of the JUNO detector.
\item In case of deviations from the average radon levels, LS batches of 3 to 5~tons were inserted into the OSIRIS system. By sourcing LS from different stages in the filling and purification chain, it was possible to quickly identify subsystems contributing additional radon and take efficient countermeasures.
\end{itemize}

The LS filling phase of JUNO CD was completed on 22~August~2025.

\subsection{LS transparency}

Regarding optical requirements, an excellent transparency and an attenuation length exceeding 20~m at a 430~nm photon wavelength are essential to achieve good energy resolution in JUNO~\cite{Zhang:LSrefractiveindex:2024,Beretta:LSfluorescence:2025}.

\begin{figure}[htbp]
\centering
\includegraphics[width=0.75\columnwidth]{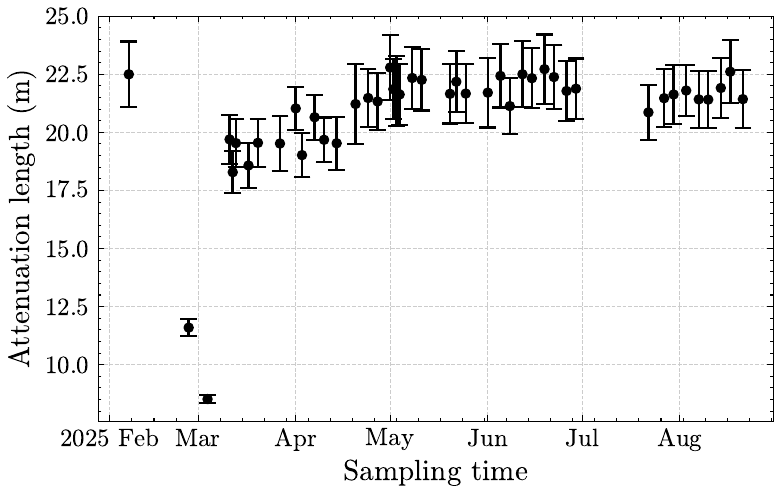}
  \caption{\label{fig:LSAttenuation}The LS attenuation length was measured throughout the six-month filling phase at a photon wavelength of 430 nm. Approximately $500~\mbox{m}^3$ of low-quality LS were identified, primarily due to yellowish PPO. The average attenuation length of all LS batches reached 20.6 m.}
\end{figure}

As previously described, a sequence of five purification plants have been
realized on-site aiming to reach the desired requirements, both for radiopurity and optical properties of the scintillator.
LS samples were regularly taken from various positions of the sequence for the attenuation length measurements using the method described in~\cite{Gao:2013pua,Yin:2020elg}.
As can be seen from Figure~\ref{fig:LSAttenuation},
although approximately 500~m$^3$ of low-quality LS were produced initially due to yellowish PPO and acid washing of the master solutions, the average attenuation length eventually reached 20.6 m at a photon wavelength of 430 nm.

\subsection{LS radio-purity}
A preliminary evaluation of the concentration of natural contaminants, $^{238}$U and $^{232}$Th, in the JUNO LS can be performed through the identification of fast decay coincidences, characteristic of both decay chains. The $^{214}$Bi-$^{214}$Po $\beta$-$\alpha$ coincidence, with an average time delay of $\Delta t = 237~\mu\mathrm{s}$, and the $^{212}$Bi-$^{212}$Po $\beta$-$\alpha$ coincidence, with an average time delay of $\Delta t = 443~\mathrm{ns}$, are powerful decay signatures that enable the determination of the progenitor concentrations, under the assumption of secular equilibrium. In this case, the assumption is not fully valid during the initial $\sim$15 days after detector filling; however, for the purpose of this work, only the average concentrations over the first two months of data taking, 30~August to 31~October~2025, are reported.
All details of the analysis, including radionuclide time evolution, event selection, and cut efficiencies, will be presented in a dedicated paper. For the present study, we select a fiducial detector volume of slightly less than 16~kton by imposing a radial cut {\marka$R < 16.5~\mathrm{m}$} and an additional cut on the vertical coordinate {\marka$|z| < 15.5~\mathrm{m}$}. Within this volume, we estimate a $^{238}$U contamination of {\marka$(7.5 \pm 0.9)\times10^{-17}~\mathrm{g/g}$} and a $^{232}$Th contamination of {\marka$(8.2 \pm 0.7)\times10^{-17}~\mathrm{g/g}$}. Both results for $^{238}\mbox{U}$ and $^{232}\mbox{Th}$ are
one order of magnitude better than JUNO requirements for the
NMO~\cite{JUNO:NMOsensitivity:2025} analysis,
and are fully compliant with the solar neutrino analysis requirements~\cite{JUNO:solar:2023}.

We have also estimated the amount of $^{210}$Po uniformly distributed within the LS. As observed in previous experiments, such as Borexino, an out-of-equilibrium concentration of this isotope is present and gradually decreases according to its 138.4-day half life. The estimated $^{210}$Po rate at the end of detector filling (late August~2025) is approximately {\marka$5\times10^{4}~\mathrm{cpd/kton}$}, while the average rate during the data-taking period from {\marka30~August~2025 to 31~October~2025} is {\marka$(4.3 \pm 0.3)\times10^{4}~\mathrm{cpd/kton}$}, within the previously defined fiducial volume ({\marka$R < 16.5~\mathrm{m}$ and $|z| < 15.5~\mathrm{m}$}).

\section{JUNO commissioning and data processing}

The JUNO experiment utilizes a full PMT waveform readout system to achieve unprecedented energy resolution, enabling precise reconstruction of charge and time. Moreover, to maximize the readout capability during supernova burst events within 3 kpc, time and charge data computed by the electronics are read out on a triggerless basis. The design of the electronics, trigger, and data acquisition systems was driven by physics requirements, and their performance was rigorously validated during the commissioning phase. This section provides a brief description of the online and offline systems' design and implementation, as well as the detector calibration strategy.

\subsection{Large PMT readout electronics and trigger}
The large PMT readout electronics is composed of two blocks:
the front-end (FE) electronics, located at few meters from the PMTs,
and the back-end (BE) and trigger electronics which is located in the two
electronics rooms in the JUNO experimental hall.
The FE electronics has been installed underwater on the same SS supporting PMTs, inside a stainless steel box called UWBox.
The installation of the UWBoxes proceeded in parallel with the construction of the JUNO detector. In total, 5878 UWBox are connected to the CD PMTs, while 803 UWboxes read out the WCD PMTs.
In addition, 116 UWboxes serve the remaining 20-inch PMTs installed on the walls of the water pool.
Apart from very few cases, a set of three PMTs is connected to one UWbox
through a $50~\Omega$, coaxial cable.
Each UWbox reads out, independently from each other, three PMTs and contains three High Voltage Units (HVU) and one Global Control Unit (GCU).

The GCU incorporates a Xilinx Kintex-7 FPGA (XC7K325T) which performs all the digital
signal processing and interacts with the Data Acquisition (DAQ) and
Slow Control (DCS) systems.
Digitalized large PMT waveforms are triggered in the FPGA via a Continuous Over Threshold Integral method, generating hit signals for global trigger and pairs of charge and time.
Besides the local memory available in the FPGA, a 2 GB DDR3 (Double Data Rate Type 3) memory is available to provide a larger buffer for collected waveforms and triggerless charge-time pairs in case of arrival of neutrinos bursts from the explosion of a Supernova.
The connection between FE and  BE/Trigger electronics is realized
thanks to a synchronous link running on a CAT6 commercially available cable; in addition, a so-called asynchronous link running through a CAT5 commercially available cable, connects the GCUs to the DAQ and DCS.
Finally, a low resistivity power cable is used to bring a 48~V power voltage to the GCUs.
These cables are embedded inside a stainless steel bellow which is welded to the UWBox on one side, and connected to the electronics room equipment, above water, on the other side.

The synchronous link provides a deterministic, low-latency, and bidirectional communication channel between the GCU and the BE electronics located above water.
It plays a critical role in the timing and trigger distribution system and also serves as the backbone for precise timing synchronization based
on the IEEE 1588-2008 Precision Time Protocol (PTP)~\cite{FPGA:1588}, enabling
nanosecond-level alignment between FE and BE components.

The Back-End Card (BEC) is responsible for direct communication with the GCUs installed underwater.
The basic encode/decode protocol between BEC and GCU is a modified
Trigger Timing and Control (TTC) link protocol~\cite{Aloisio:TTC}.
The BEC has been designed to root the incoming trigger request signal and
the TTC commands from the GCU and to distribute a 62.5 MHz clock signal to each GCU, and the trigger validation signals from the trigger electronics.
A Reorganize and Multiplex Unit (RMU)~\cite{Aloisio:RMU}, hosts three mezzanine cards and is responsible for communication between 21 BECs and the Central Trigger Unit (CTU).

A global clock and triggering system is crucial to ensure high-speed and efficient  readout of the waveforms for a large number of channels.
The global trigger (also called multiplicity trigger)
scheme relies on information from all PMTs to form a unified trigger decision that initiates the readout process for all instrumented PMTs.
In each GCU, the digitized waveforms are processed in the FPGA and, if the
signal goes above a threshold of 5 times the electronic noise of a channel, a logical `hit' signal is generated.
The `hit' signal goes back to zero, when the waveform goes below
the same threshold. A trigger primitive is computed by summing the `hit' signals over a maximum of 3 PMTs connected to the GCU.
The sum is computed every 16~ns (1 Clock cycle).
This logical signal is sent to BEC through the synchronous connection.
Inside each BEC, all trigger validation signals coming from a maximum of
48 GCUs (i.e. maximum 144 PMTs) are summed and sent to the CTU, through the RMU.
The CTU computes a global trigger validation by adding all the received signals and compares them to a single threshold.
The sum is computed in a 304~ns time window and in case a threshold of 350 PMTs is reached, a trigger validation signal
is broadcasted to all the GCUs (through the RMUs and BECs).
A trigger timestamp is transmitted to all GCUs together with the trigger validation signal.
In addition to the multiplicity trigger, a periodic trigger at a 50~Hz rate
is issued during normal runs to study the PMT dark noise.

\subsection{Small PMT readout electronics}
The readout electronics~\cite{JUNO:2025dfn} for the 25,587 3-inch photomultipliers (small PMTs) follows a similar two-tier architecture as the large PMT system, with front-end (FE) electronics located 5 to 10 meters from the PMTs and back-end electronics installed in the surface electronics rooms of the experimental hall. The FE electronics are housed underwater in cylindrical stainless-steel enclosures, the UWBs, mounted on the detector's stainless-steel structure supporting the acrylic vessel and PMTs. Each UWB serves 128 PMTs, organized as eight groups of sixteen photomultipliers selected to have similar gain and glass thickness so that a common high voltage can be applied at a common depth in the water pool.
Each group of sixteen PMTs is powered through a High Voltage Unit~(HVU), and sixteen HVUs are installed per UWB to provide redundancy. The HVUs are integrated on two large 64-channel High-Voltage Splitter (HVS) boards~\cite{WALKER2026171022} that distribute the high voltage and simultaneously decouple the PMT analog signals from the same coaxial cables. The 128 coaxial cables connecting the PMTs are interfaced to the UWB through eight underwater feed-through connectors, each handling sixteen channels. The decoupled signals are transmitted to the ABC front-end readout board, which digitizes and packages charge and time information via eight 16-channel CATIROC ASICs \cite{JUNO:2020orn} controlled by a Kintex-7 FPGA. The associated GCU control board provides power regulation, slow control, synchronization, and data transmission to the (back-end) DAQ. The complete system achieves low noise levels of 0.04 PE well below (its self-trigger level) the targeted 0.33 PE trigger threshold, a crosstalk below 0.4\% across the channels, and a bandwidth of 57 MB/s capable of handling high-rate scenarios.

\subsection{Large PMTs data acquisition and Online Event Classification}
All the large PMT GCUs asynchronous links,
are connected to dedicated network switches,
and from there through optical fiber to the DAQ servers.
The JUNO DAQ~\cite{JUNO:DAQ} is a distributed system which is responsible for
acquiring and processes the GCU data stream and assemble it with the trigger information coming from the CTU.
For each trigger, a customizable readout window of $1~\mu\mbox{s}$ (1008 ns, to be precise) is extracted from the digitized stream and packeted in a 2032 bytes waveform packet containing the signal samples and metadata (PMT identification code, timestamp, and other data).
The DAQ sorts the single PMT waveforms according to the timestamp and performs the event building.
Data coming from the CD and WCD are processed in the same way, but in separate streams.
At a trigger rate of ~500~Hz, a large amount of data (up to 20~GB/s) is generated.

The Online Event Classification (OEC) system, which aims to identify reactor and atmospheric neutrino events to save full waveforms, to identify events only saving waveforms for fired PMTs, to tag event of less interest to save reconstructed charge and time information only, is therefore performed on the basis of physics considerations.
The system reconstructs in real-time PMT waveforms and position and energy of each event.
The reconstructed quantities for each event are written to a buffer called OECevt that contains vertex, energy, track and preliminary tag information (i.e., low, medium, high energy).
The OECevts are ayalyzed in an individual High Event Classification (HEC) node which is capable of identifying spatial and temporal correlations and CD-WP correlations. According to high-level tags (e.g. Inverse Beta Decay's prompt-delayed, primary muon, spallation neutron), OEC decides whether to save or discard the waveforms collected by the DAQ for a given event. In this way, 90~MB/s raw data are saved to disk and transferred to offline computing centres.

Raw data quality and detector performance are monitored in real time by the DAQ and the OEC systems. The OEC Monitor is designed to track detector performance and data quality in real-time by leveraging event reconstruction and classification results. It processes both channel-level and event-level information to generate a suite of configurable metrics and histograms. These metrics and histograms are customizable via JSON settings and are displayed through the DAQ web interface, enabling shifters to promptly assess detector status. A particularly critical function of the OEC Monitor is its capability to monitor $^{222}$Rn rate in CD during liquid scintillator filling, calibration source deployment, or other activities with risks of radon leakage. This rapid detection assisted shifters in locating the issue and taking timely actions, such as halting LS filling, thereby helping to maintain the low background conditions essential for achieving the experiment's physics goals.


\subsection{The offline and computing framework}

The JUNO offline software system (JUNOSW), has been built upon the SNiPER (Software for Non-collider Physics Experiments) framework~\cite{Zou:2015ioy}.
JUNOSW is structured around several core components to ensure efficient and scalable data processing:
\begin{itemize}
\item Algorithms, which perform event-level operations such as calibration and reconstruction;
\item Services, which provide shared functionalities such as geometry management~\cite{Li:2018fny} and input/output (I/O) handling;
\item Tasks, which orchestrate the execution flow and mediate interactions among algorithms and services;
\item the Event Data Model (EDM)~\cite{JUNO:EDM} which is built on ROOT~\cite{brun1997root} and it facilitates efficient data storage and retrieval with support for scheme evolution;
\item and the visualization tools~\cite{You:2017zfr,Zhu:2018mzu}, which combine the detector geometry~\cite{Song:2025pnt} and event data to display detector and event information for simulation, reconstruction and analysis.
\end{itemize}

Once raw data arrive at the IHEP data centre, they are reprocessed and converted into ROOT-based RAW (RTRaw) data using JUNO EDM.
During the calibration stage, RTRaw data undergo
waveform reconstruction and channel-to-channel variation corrections to produce calibrated event records.
This is followed by the reconstruction stage, which extracts key physical
observables, such as the total deposited energy, the interaction vertex
coordinates point-like events, and the direction of track-like events.
The outputs from both stages, calibrated data and reconstructed results
are consolidated into the Event Summary Data (ESD) format,
which serves as the primary input for subsequent physics analyses.

JUNO collaboration owns a Distributed Computing Infrastructure (DCI),
an implementation of a grid computing system based on Worldwide LHC Computing Grid (WLCG).
JUNO uses DIRAC~\cite{Tsaregorodtsev:2008zz} as main JUNO DCI core, and then a set of services installed in
JUNO DCI sites: IHEP Computing Centre in China, CC-IN2P3 in France,
INFN CNAF in Italy, and JINR in Russia.
In this model, IHEP serves as the Tier-0 data centre, while other data centres in Europe function as the Tier-1.
raw-data in byte stream format is transferred from the JUNO on-site to Tier-0 via a dedicated network (100 Gbps between IHEP and Europe, 10 Gbps between INFN CNAF and JINR).
Upon arrival, the data is registered in a DIRAC-based file catalogue and subsequently distributed to Tier-1 centres.
The raw-data file is also archived in a tape library with two redundant copies.
\subsection{Calibration strategy}

\begin{figure}[htbp]
\centering
  \includegraphics[width=0.675\columnwidth]{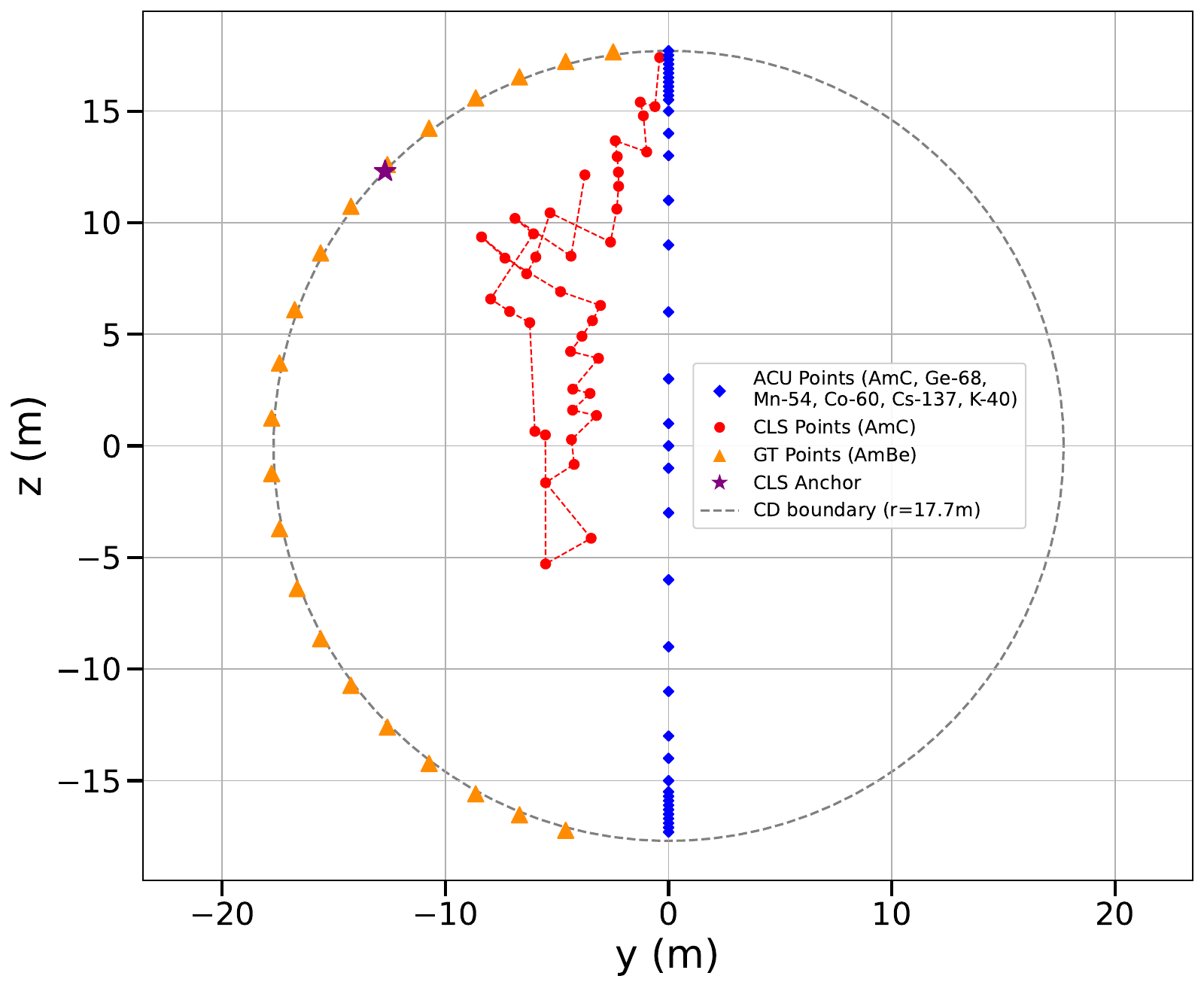}
  \caption{\label{fig:calib:points} Sources positions during
  the intensive calibration runs after LS filling finished. The blue points show the positioning along the z-axis of the source with the ACU, the orange points show the GT positioning along the acrylic vessel surface, while the red points show an example of points taken on the z-y plane with the CLS.}
\end{figure}

The ACU has been designed as a primary tool to precisely calibrate the
energy scale of the detector, to align the PMTs timing,
and partially monitor the position-dependent energy scale variations.
The design~\cite{calib:ACU} foresees four independent spools mounted on a turntable. Each spool is capable to unwind and deliver the source via gravity through the central chimney of the CD,
with positioning precision along the z-axis better than 1~cm.
During the routine calibration runs, three sources are regularly deployed :
a neutron source, a gamma source and a pulsed UV laser source carried by an optical fibre with a diffuser ball attached to the
end~\cite{calib:laser,Takenaka:2024ctk}. The fourth spool can either carry a radioactive
source, or a temperature sensor, or a floater to help monitor the interface of LS and water.

By means of CLS, sources are moved on a z-y plane. The relative
precision on the source positioning has been measured to be better than
3~cm. The GT allows to deploy a source along the acrylic vessel surface at two specific azimuthal angles, $\varphi = 123.4^\circ$ and $\varphi = 267.4^\circ$.
It allows to verify the response of the detector at the boundary between the LS and the PMT water buffer. During the GT calibration runs, an Am-Be source is used as a proxy of prompt and delayed signal.

The operation of the CLS system was incompatible with the LS filling process due to a conflict in the top chimney, preventing its use for calibration during this period. However, upon the LS reaching the bottom level, an Am-Be neutron source was deployed by the GT system at positions z = $-$16 m and $-$17.2 m. A comparison between the monitored neutron rate as a function of LS level and the corresponding simulations allowed for a precise prediction of the LS exchange end time, with an uncertainty of merely 12 minutes. This corresponds to a precision of $6\times10^{-5}$ relative to the total six-months exchange duration. Subsequent to the filling of the CD, a comprehensive calibration campaign was conducted utilizing the ACU, GT, and CLS systems, as depicted in Figure~\ref{fig:calib:points}.

During the stable physics data taking period, UV laser runs were taken weekly or after hardware changes. The Am-C neutron source was bi-weekly deployed. Another comprehensive calibration with more CLS positions is under preparation.

\section{Detector performance results}

Accurate energy measurement in LS detectors is fundamentally dependent on the precise reconstruction of charge and time from PMT waveforms. This section details the foundational steps taken to achieve this in JUNO. We describe the careful tuning of the PMT high voltages and the calibration of their gains and time offsets using a laser source. The results of these procedures are presented herein.

\subsection{Large PMT waveform reconstruction}
\begin{figure}[htbp]
\centering
  \includegraphics[width=0.586\columnwidth]{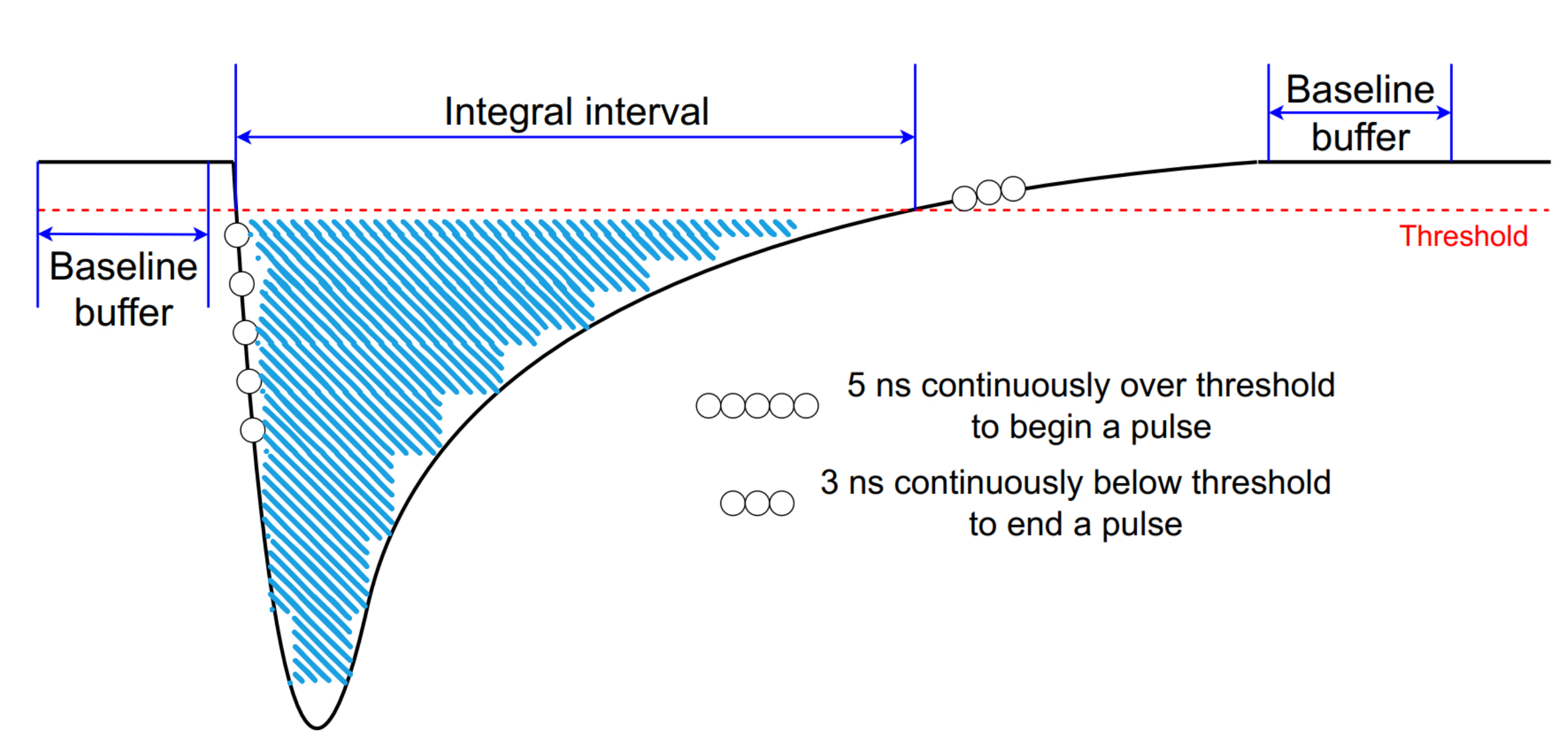}
  \includegraphics[width=0.378\columnwidth]{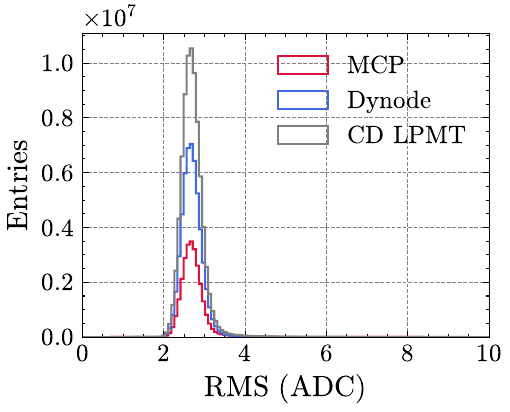}
  \caption{\label{fig:COTI}
  Left: Principle of the Continuous Over Threshold Integral (COTI) method for the waveform reconstruction. Further details are in the text.
  Right: Pedestal RMS of all CD large PMT channels. The average noise is 2.6 ADC count~(0.16~mV), equivalent to about 0.05~PE. in amplitude for PMTs running at 10$^{7}$ gain.}
\end{figure}

The JUNO analysis chain starts from waveform reconstruction of 20-inch PMTs.
Two different waveform reconstruction algorithms have been developed: COTI (Continuous Over-Threshold Integral) and a deconvolution-based method.
Due to its simplicity and comparable robustness to the deconvolution method, the COTI algorithm has been adopted as baseline.
The basic principle of COTI involves the following steps, also shown in Figure~\ref{fig:COTI}:
\begin{itemize}
    \item Calculate the baseline by averaging the initial 24~ns segment of the digitized PMT waveform.
    \item Move to the next 8~ns time window, and search for five consecutive points that surpass the pre-defined threshold relative to baseline.
    \item If such a sequence is detected, it indicates the presence of a pulse. The first over-threshold point will be identified as the start of the pulse, and the baseline will be kept unchanged. Otherwise, the baseline calculation window will shift forward by 8~ns and be updated.
    \item Upon identifying a pulse, COTI searches the subsequent 8~ns window for three consecutive points falling below the threshold. The first of them will be designated as the pulse end time.
    \item Finally, the pulse charge is computed by numerically integrating the waveform between the start and end points.
\end{itemize}

The electronics grounding has been carefully designed and built. The average pedestal RMS is about 2.6~ADC count~(0.16~mV), equivalent to about 0.05~PE in amplitude for PMTs running at 10$^{7}$ gain.
This excellent low noise level is the result of a successful grounding scheme and isolation between the dirty and clean grounds. It allowed the PMTs running at lower gains while keeping a reasonable good trigger efficiency.
After considering the PMT waveform shape, equivalent threshold is about $100~\mbox{ADC}\cdot\mbox{ns}$ and 0.25~PE for JUNO large PMTs running at (6 to 7)$\times10^6$ gain.

\subsection{Large PMT running status}

A total number of 20,348 20-inch PMTs have been installed and
operated in JUNO since the LS filling phase.
%
Table~\ref{tab:pmt:summary} shows the total number of 20-inch PMTs
installed in CD and WCD, respectively.
A total of 22 PMTs ~(0.11\%) have been found dead until November~2025, possibly due to water leakage, or unstable operation of the
HV modules.
About 2\% to 3\% of the operational PMTs are found be to flashing, with a typical phenomenon of a sudden increase of counting rate from the flasher PMT.
As a result, the peak rate could reach more than 1~MHz.
The potential origin of the flashing process could come from discharge of spacers or other components in the PMTs~\cite{hkk::flasher:private-communication,nnvt::flasher:private-communication}.
It is expected that a large fraction of flashing PMTs can be recovered in the near future.
\begin{table*}[ht]
\centering
\begin{tabular}{|l|rr|rr||r|}
\cline{1-5}
    & \multicolumn{2}{c|}{\textbf{CD}} & \multicolumn{2}{c|}{\textbf{WCD}}\\
    \cline{2-6}
    & \textbf{Hamamatsu} & \textbf{NNVT} &
       \multicolumn{2}{c|}{\textbf{NNVT}} & \\
    & \textbf{R12860-50} & \textbf{GDB6201} &
       \multicolumn{2}{c|}{\textbf{GDB6201}} & \textbf{Total}\\ \hline
Installed         &  4939 & 12657 &  2404 &   348 & 20348\\ \hline\hline
In operation &  4936 & 12638 &  2404 &   348 & 20326\\
Dead              &     3 &    19 &     0 &    0 &    22\\
\hline
\end{tabular}
\caption{Summary of the running status of the large PMTs installed in the CD and WCD. Dead PMTs are primarily due to HV trip.}
\label{tab:pmt:summary}
\end{table*}

\begin{figure}[htbp]
\centering
  \includegraphics[width=0.495\columnwidth]{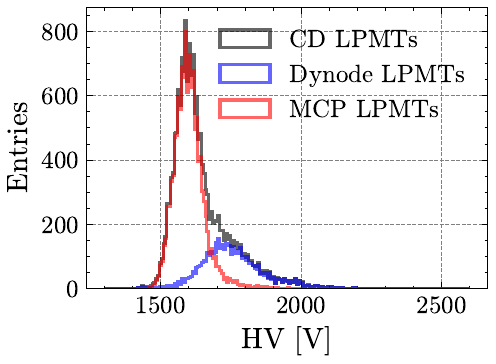}
  \includegraphics[width=0.495\columnwidth]{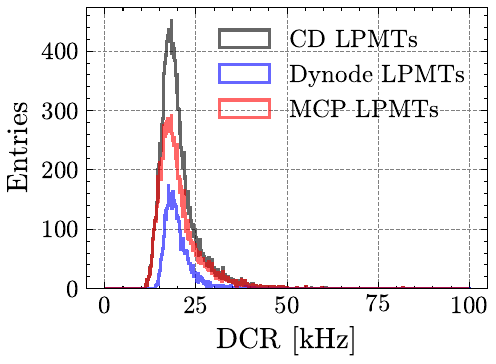}
  \caption{\label{fig:lpmt:dcr_hv} Left: High Voltage distribution of the large PMTs at an operational gain of $0.65\times 10^{7}$ and $0.72\times10^{7}$ for dynode-PMTs and MCP-PMTs, respectively.
  Right: Large PMTs DCR distribution.}
\end{figure}

The operational high voltages of the PMTs (see left plot of Figure~\ref{fig:lpmt:dcr_hv}) were established based on a comprehensive analysis of the PMT detection efficiency, single-photoelectron charge spectrum, single-channel trigger threshold, and flashing probability. Contributions from the dynode and MCP-PMTs are presented separately.

The dark count rate (DCR) is a critical parameter in PMT operation, monitored through both online hit counting in the electronics and offline analysis of periodic triggers. The right plot of Figure~\ref{fig:lpmt:dcr_hv} shows the offline-measured DCR distribution during a typical physics run.
The average DCR values are 20.6 kHz and 22.7 kHz for dynode and MCP-PMTs, respectively. While a direct comparison of these values with the mass testing results in~\cite{JUNO:PMTmassTesting:2022} is challenging due to differing operational conditions, such as ambient temperature, PMT gain, trigger thresholds, and electronics noise levels, a decrease of the average DCR with time was observed.

\subsection{Large PMT gains}

To characterize the single-photoelectron (SPE) charge distribution, two different gain definitions are introduced: the peak gain ($G_p$) which corresponds to the peak position of the SPE charge distribution, and the mean gain ($G_m$), which represents the average value of the SPE charge distribution.
The PMT gain is extracted by fitting the charge spectra obtained from calibration or periodic trigger data. For calibration, a stable light intensity ensures that the Number of Photo-Electrons (NPE) detected by each PMT follows a Poisson distribution with mean occupancy $\mu$. This is achieved using either a radioactive or laser source in the CD centre: for the former, events are selected by vertex and energy cuts around the source peak; for the latter, an external trigger is provided by a monitor PMT illuminated by a split laser beam.
For periodic trigger data, dominated by PMT dark noise, a clean sample of SPE waveforms is obtained by selecting single-pulse events. Minor contamination from scintillator or detector radioactivity introduces a small multiple-PE fraction, which is neglected in the fit. In this work, laser data are used to determine the PMT gain precisely, while periodic trigger and laser calibration data are employed to extract the DCR and monitor gain stability.

\begin{figure}[hbt]
    \centering
    \includegraphics[width=0.45\linewidth]{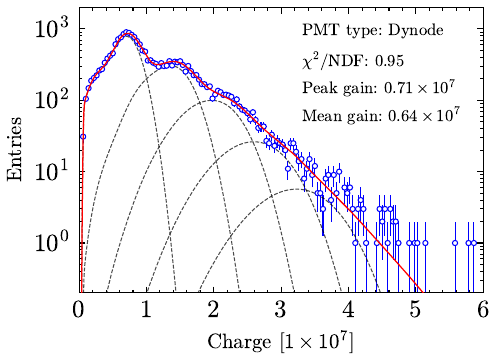}
    \includegraphics[width=0.45\linewidth]{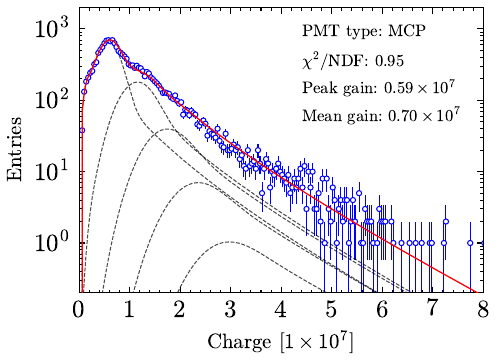}
    \caption{Representative examples of the PMT gain calibration fits. The left and right figures show the results for a dynode- and an MCP-PMT, respectively.
    The solid curves represent the best fit to the charge spectra, with the individual contributions from up to 5 PEs shown as dashed lines.}
    \label{fig:gain_calib_fit}
\end{figure}
Figure~\ref{fig:gain_calib_fit} shows two typical charge spectra from laser calibration data, one from a
dynode-PMT (left) and the other from an MCP-PMT (right).
It's worth noting that both charge spectra exhibit distinct structures: the dynode-PMT shows a small shoulder in the low-charge region, whereas the MCP-PMT features a long tail in the high-charge region.
To account for these characteristics, two different SPE charge response models have been developed: a double Gaussian model for the HPK dynode-PMT and a recursive model for the MCP-PMT. The measured charge spectrum from the laser source incorporates not only the SPE component but also multiple-PE components.
The latter can be expressed as the $n$-times convolution of the SPE charge spectrum. Due to complicated SPE charge response model, it's challenging to derive the analytical form for the multiple-PE spectrum, we instead adopt a fast numerical method based on Fourier transforms, as proposed in~\cite{Kalousis:2019lua}.
The red curves in Figure~\ref{fig:gain_calib_fit} represent the best fits to the charge spectra, with the dashed lines indicating the contributions from up to 5 PEs.
\begin{figure}[h]
    \centering
    \includegraphics[width=0.9\linewidth]{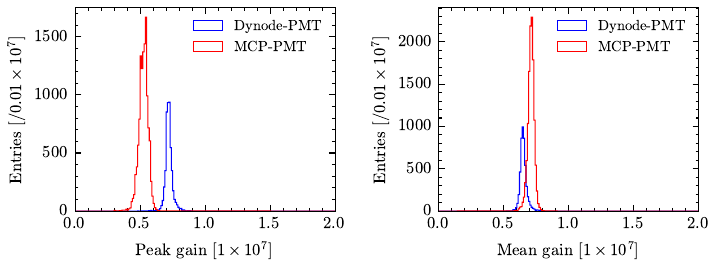}
    \caption{(Left) Peak gain, $G_p$, and (Right) mean gain, $G_m$, distributions for both dynode-PMT (blue) and MCP-PMT (red).}
    \label{fig:gain_distribution}
\end{figure}

\begin{figure}
    \centering
    \includegraphics[width=0.9\linewidth]{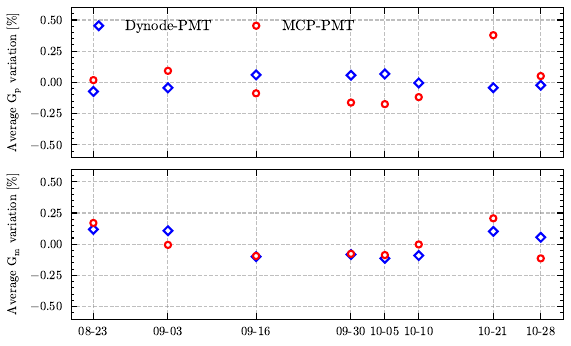}
    \caption{Average peak gain and mean gain evolution in time since the beginning of the JUNO science runs. Each point is determined from a new laser calibration run.}
    \label{fig:gain_stability}
\end{figure}

Figure~\ref{fig:gain_distribution} shows the peak gain (left) and mean gain (right) distributions for the two types of large PMTs. For dynode PMTs, the average peak gain is $G_p = 0.72\cdot 10^{7}$ and while the average mean gain is $G_m = 0.65\cdot 10^{7}$. For MCP-PMT, the average peak gain is $G_p = 0.53\cdot 10^{7}$ and the average mean gain is $G_m = 0.72\cdot 10^{7}$.
The mean gain is considered a more physical quantity, and dividing the charge of a pulse by the mean gain yields an unbiased estimate of NPE.
Gain stability in time is monitored using laser calibration data, and the result are shown in Figure~\ref{fig:gain_stability}.
Both dynode-PMTs and MCP-PMTs demonstrate a remarkable gain stability throughout the first two month of science runs data.

\subsection{Large PMT timing}
PMT time synchronization is a very important ingredient for event vertex
reconstruction and data analysis.
In JUNO the synchronous link operates across long-distance CAT6 cables (up to 100~m), and it involves distinct and asynchronous clock domains on the transmitter BEC and receiver GCU sides.
Although the clocks are frequency-synchronized, using a modified dedicated version of the IEEE 1588 protocol~\cite{FPGA:1588}, they may exhibit an unknown and random time-offset that must be recalibrated with an ad-hoc procedure.
To align all the GCUs time offsets, special calibration runs with a laser source are performed. The laser~\cite{PiLas} is a picosecond laser with a tunable pulse width between 15~ps and 50~ps and with optical peak powers between 50~mW and 1.5~W.
It is possible to operate it with high repetition rates with very low jitters.
Two optical heads are employed in JUNO with wavelength of $420~\mbox{nm}$ and $266~\mbox{nm}$: the JUNO LS is transparent to the former frequency photons, while the latter wavelength photons are completely absorbed by JUNO LS and create scintillation photons that propagate in the LS towards the PMTs.

The setups used for the laser calibration runs are the following:
\begin{itemize}
\item the laser head is connected through an optical fiber to a half-reflecting mirror that splits the laser beam.
\item one part of the laser is sent to the JUNO detector inside an optical
fiber connected to an optical diffuser ball (made of PTFE) which is deployed with the ACU along the central axis of the CD;
\item the other part of the beam reaches the photo-cathode of a monitor PMT
(model R8520-406, from HPK) which is used to measure the $t_0$ of the events, thanks to its very small transit time spread (less than 1~ns).
This values are computed on a single event basis and stored in the raw-data for timing corrections.
\end{itemize}

\begin{figure}[htbp]
\centering
    \includegraphics[width=0.475\columnwidth]{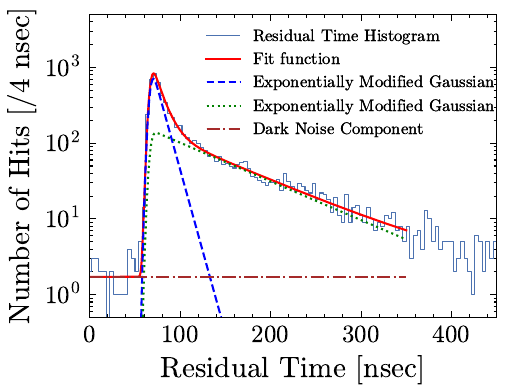}
    \includegraphics[width=0.475\columnwidth]{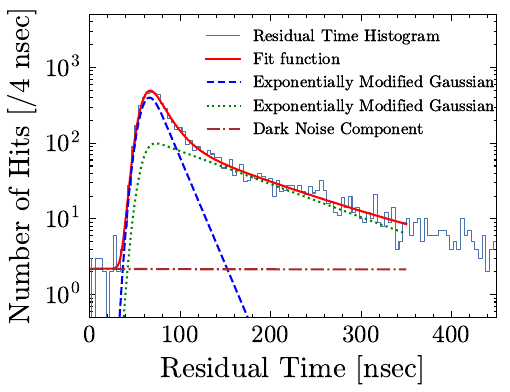}
  \caption{\label{fig:timing:pmt-example}Residual time distribution for
  a typical dynode-PMT (left) and MCP-PMT (right). The curves on the plot
  represent a fit to the data with different components. A description is given in the text.}
\end{figure}

The plots of Figure~\ref{fig:timing:pmt-example} show a typical comparison example of the residual time distribution, for a dynode-PMT (left) and for a MCP-PMT (right), defined as:
\begin{equation}
    t = t_{\mbox{first-hit}} - t_{\mbox{TOF}} - t_{\mbox{ref-PMT}}~\mbox{,}
\end{equation}
where $t_{\mbox{first-hit}}$ is the time of the first photon giving a signal in any CD PMT, $t_{\mbox{TOF}}$ is the time needed by the optical photons to reach the PMTs' photocathode and $t_{\mbox{ref-PMT}}$ is the time measured by the monitor PMT.
The residual timing distribution observed by each PMT in the LS phase is fitted with two exponentially modified Gaussians, each defined as the convolution of a Gaussian and an exponential function, both superimposed on a dark-noise component to extract the peak position and the width of the distribution:
\begin{align}
f(t) &= \frac{\alpha}{2\tau_{1}}\exp\left(\frac{2\mu+\sigma^{2}/\tau_{1}-2t}{2\tau_{1}}\right){\rm erfc}\left(\frac{\mu+\sigma^{2}/\tau_{1}-t}{\sqrt{2}\sigma}\right) \notag \\
&+ \frac{(1-\alpha)}{2\tau_{2}}\exp\left(\frac{2\mu+\sigma^{2}/\tau_{2}-2t}{2\tau_{2}}\right){\rm erfc}\left(\frac{\mu+\sigma^{2}/\tau_{2}-t}{\sqrt{2}\sigma}\right), \label{eq:signaltime}
\end{align}
In this model, the Gaussian term ($\mu$ and $\sigma$) describes the PMT time response, while the exponential term represents the LS light-emission time profile. The parameters $\tau_{1}$ and $\tau_{2}$ correspond to the fast and slow components of the LS emission, respectively.

\begin{figure}[htbp]
\centering
  \includegraphics[width=0.475\columnwidth]{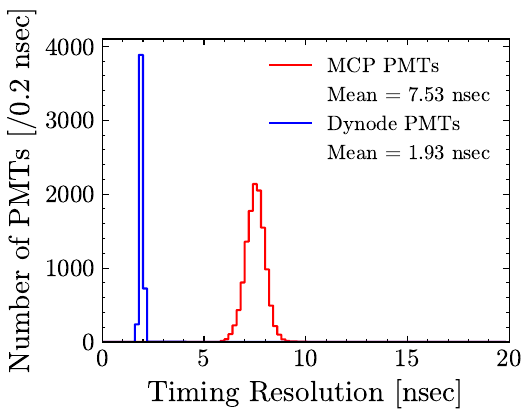}
  \includegraphics[width=0.475\columnwidth]{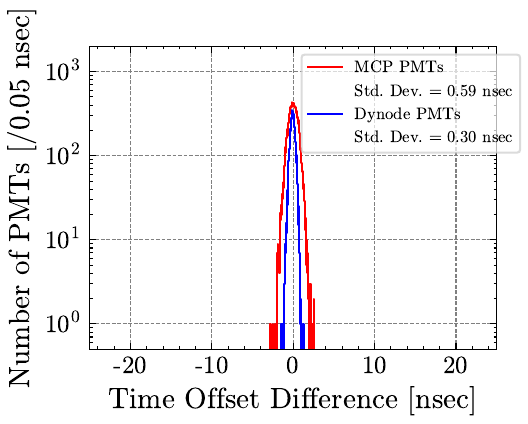}
  \caption{\label{fig:timing:reso_stability}
  Left: PMT timing resolution extracted with a laser run. The value is
  dominated by the PMT Transit Time Spread.
  Right: PMT Time offset difference measured in different runs taken on 10~October~2025 and 17~October~2025. Timing of all large PMTs kept stable in one week.}
\end{figure}

An estimation of the PMTs Transit Time Spread (TTS) can be extracted using a laser run taken during the full water phase of JUNO. One of the parameters of the model used to fit the PMT residual time is proportional to the PMT TTS.
The left plot of Figure~\ref{fig:timing:reso_stability} shows the distribution of the PMT TTS for the dynode- (blue curve) and MCP-PMTs (red curve) as measured during the JUNO water phase runs.
The right plot of Figure~\ref{fig:timing:reso_stability} shows a comparison of the PMT time offset in different runs, as measured during the science runs. The spread of the distribution show that the PMTs waveform readout can be aligned to better than 1~ns, that is the ADC sampling time.

\section{Energy calibration and reconstruction in the CD}

The detector's energy response is a key parameter in neutrino oscillation experiments. In large liquid scintillator detectors, this response is typically described in terms of three components: energy resolution, particle- and energy-dependence (non-linearity), and spatial dependence (non-uniformity). This section first introduces the light yield of the JUNO detector, then describes the methods used for vertex and energy reconstruction. The performance of these reconstruction algorithms is then presented.

\subsection{Calibration sources}
Regular calibration runs are carried out to ensure stable control of the JUNO detector's energy scale. At the beginning of the science runs, between 23 and 26~August~2025, an extensive calibration campaign was conducted to determine the overall calibration constants. Thereafter, weekly calibration runs are performed with an Am-C neutron and laser sources.

\begin{table*}[ht]
\centering
\begin{threeparttable}

\begin{tabularx}{\textwidth}{l|X|c|c>{\raggedriincludingght\arraybackslash}X}
\hline
\textbf{Source} & \textbf{Energy} & \textbf{System} & \textbf{Activity}  \\
\hline
\multicolumn{4}{c}{\textbf{Gamma Sources}} \\
\hline
$^{68}$Ge   & 0.511 $\times$ 2 MeV & ACU & 595~Bq\tnote{a}   \\
$^{137}$Cs  & 0.662 MeV & ACU & 140~Bq\tnote{b}  \\
$^{54}$Mn   & 0.835 MeV & ACU & 521~Bq\tnote{a}   \\
$^{40}$K    & 1.460 MeV & ACU & 13~Bq  \\
$^{60}$Co   & 1.173+1.333 MeV & ACU & 165~Bq \tnote{b}  \\
\hline
\multicolumn{4}{c}{\textbf{Neutron Sources}} \\
\hline
$^{241}$Am--$^{13}$C & \makecell[l]{%
neutron + 6.13 MeV ($^{16}$O$^{*}$)\\
(n,$\gamma$)p         2.223 MeV\\
(n,$\gamma$)$^{12}$C  4.94 MeV\\
(n,$\gamma$)$^{56}$Fe 7.63 MeV, etc.\\
}  & \makecell[c]{%
ACU\\CLS
}  & \makecell[c]{%
130~Bq\\ 100 Bq}  \\
\hline
$^{241}$Am--$^{9}$Be   & \makecell[l]{%
neutron + 4.43 MeV ($^{12}$C$^{*}$)\\
(n,$\gamma$)p 2.22 MeV\\
}& GT & 30~Bq   \\
\hline
\multicolumn{4}{c}{\textbf{Optical Calibration}} \\
\hline
Laser & \makecell[c]{%
Optical pulses \\ (420 nm and 266 nm) }& ACU & 50~Hz \\
\hline
\end{tabularx}
\begin{tablenotes}
\footnotesize
\item[a] Reference activity measured on 6/21/2023.
\item[b] Reference activity measured on 4/6/2021.
\end{tablenotes}
\caption{\label{tab:sourcelist}Artificial $\gamma$, neutron, and laser calibration sources operated in JUNO}
\end{threeparttable}
\end{table*}

Table~\ref{tab:sourcelist} show a list of calibration sources
 and laser used during JUNO's commissioning and first science runs;
they are grouped according to particle type
($\gamma$, n and laser photons). Most of the sources were deployed with
the ACU and used to perform scans along the CD central z-axis.
The neutron sources were operated also with the CLS and GT calibration systems.

\begin{figure}[htbp]
\centering
  \includegraphics[width=0.875\columnwidth]{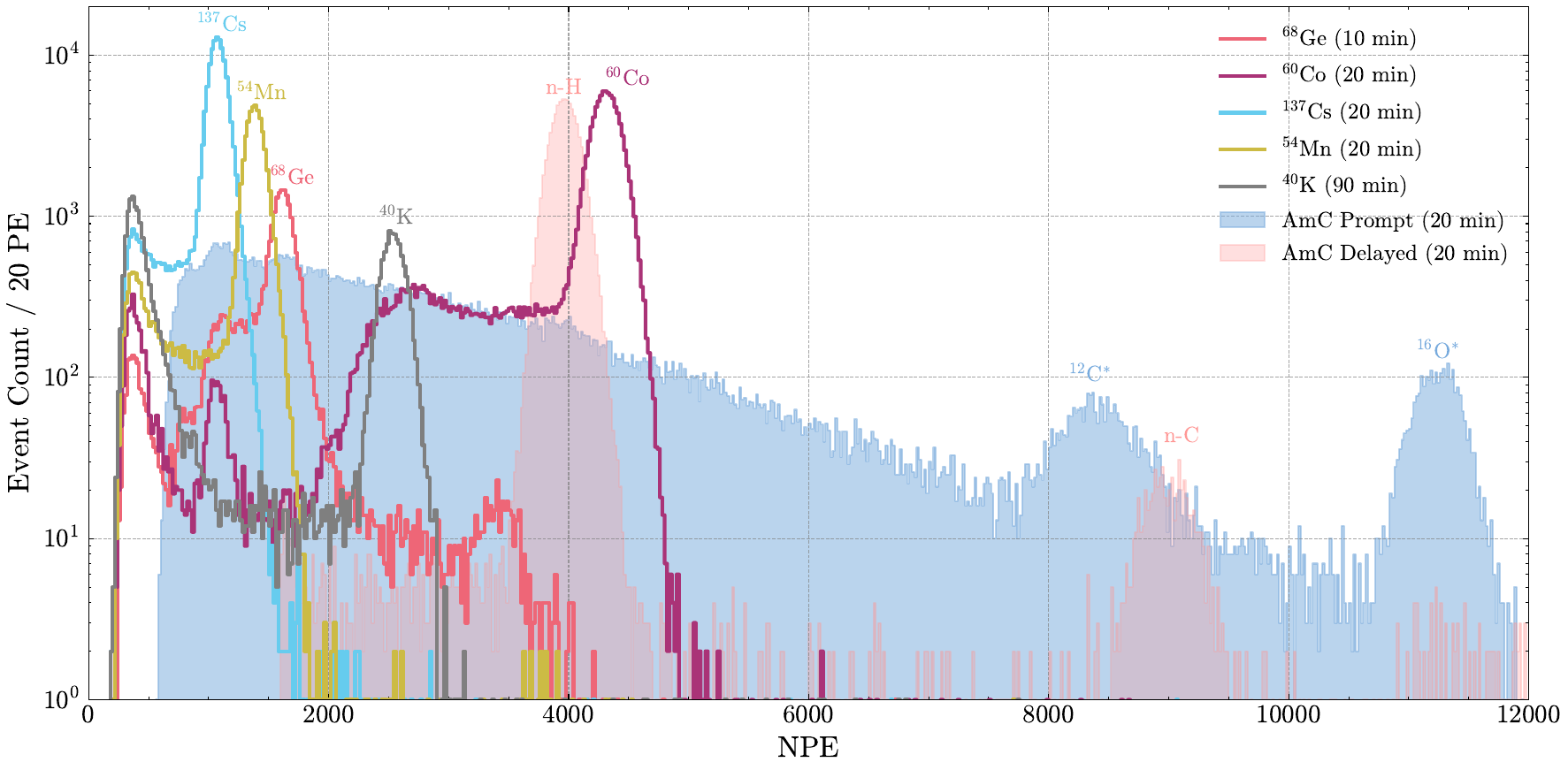}
  \caption{\label{fig:energy:totalPE} Charge distributions~(NPE) from different gamma and neutron calibration sources, after subtraction of PMT dark noise.}
\end{figure}

Figure~\ref{fig:energy:totalPE} shows the visible energy, expressed
as NPE, collected during different calibration runs.
The plot shows the position of the full energy peaks for the gamma sources (as lines) and of the prompt and delayed peaks for the Am-C source (blue and red filled areas, respectively).
As illustrated, the set of radioactive sources used in JUNO spans a wide energy range. The NPE is obtained as the sum of the reconstructed photoelectrons from all large PMTs: for each PMT, the reconstructed PE is defined as the ratio of the charge integral to the mean gain. The average contribution from the PMT dark noise, as measured using the periodic trigger for each run, has been subtracted from each event.

\begin{figure}[htbp]
\centering
  \includegraphics[width=0.485\columnwidth]{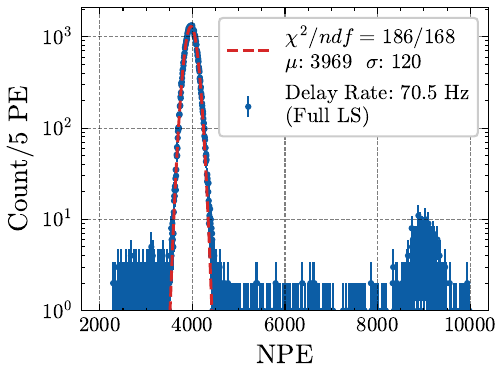}
  \includegraphics[width=0.435\columnwidth]{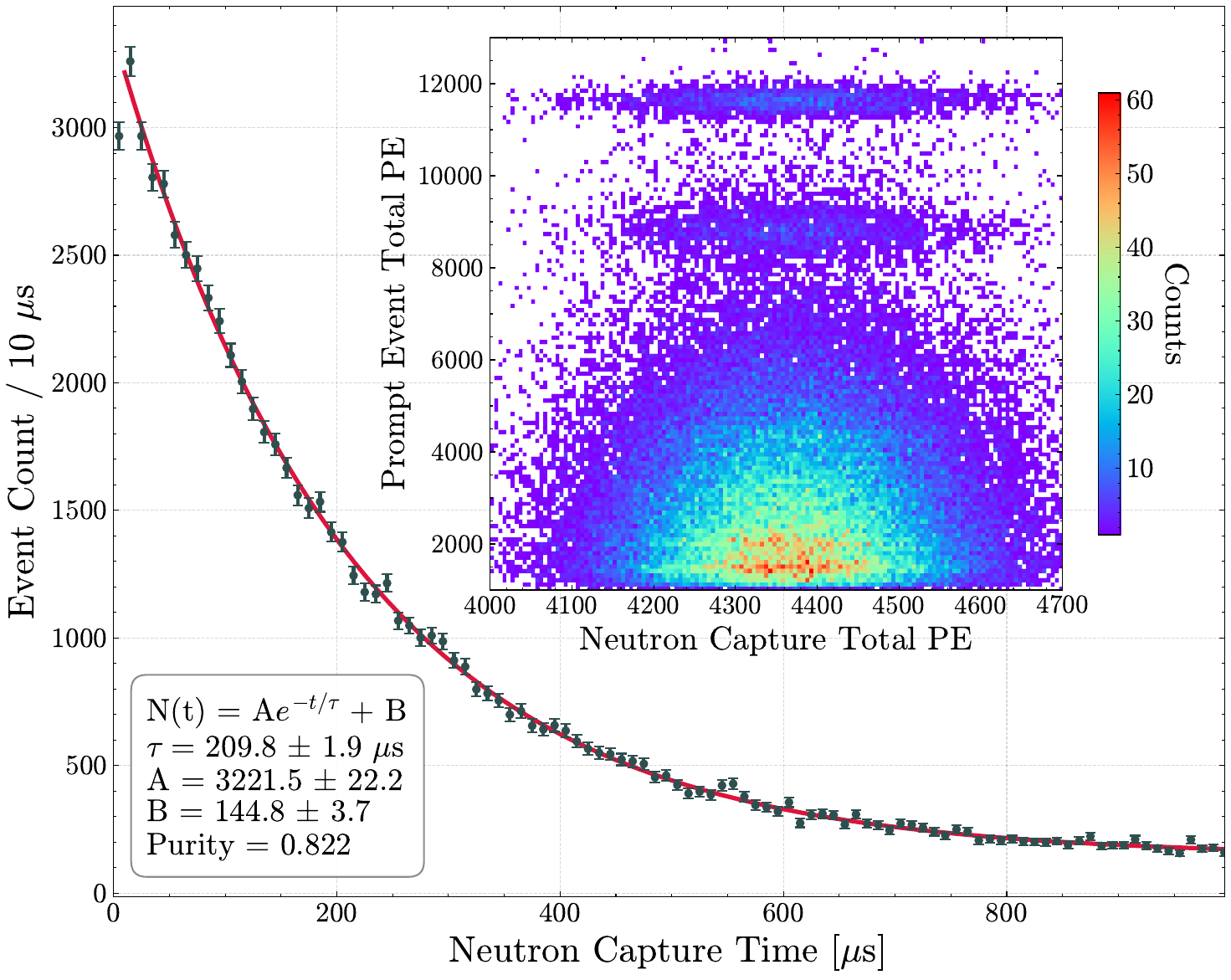}
  \caption{\label{fig:energy:AmC}Left: AmC NPE spectrum released by n capture on H (lower energy peak) or $^{12}$C (higher energy peak). Right: Time difference between the prompt signal and neutron capture. The deviation of the first point is due to a $\Delta t >$ 5~$\mu$s cut and is therefore not included in the fit. }
\end{figure}

A radioactive Am-C calibration has been chosen as neutron source in JUNO. The $\alpha$ decaying from $^{241}\mbox{Am}$ interacts with $^{13}\mbox{C}$ producing a neutron and an $^{16}\mbox{O}^{*}$ excited state.
The latter de-excites with the emission of one or two $\gamma$. The neutrons thermalize in the LS and eventually are captured by either a $^1$H or $^{12}\mbox{C}$ nucleus, resulting in characteristic $\gamma$ emissions~2.223 MeV and 4.94 MeV, respectively.
This results in a characteristic prompt/delayed coincidence event signature, similar to that produced by an Inverse Beta Decay (IBD) interaction.
The left plot of Figure~\ref{fig:energy:AmC} shows a typical NPE spectrum for neutron capture on H (left peak) and on $^{12}\mbox{C}$ (right peak).
The right plot of Figure~\ref{fig:energy:AmC} shows the time difference between the delayed (neutron capture) and the prompt, $^{16}\mbox{O}^{*} \rightarrow ^{16}\mbox{O} + \gamma$, signals.
Data exhibit a typical exponential shape with a neutron capture time, $\tau = (212.9 \pm 1.9)~\mu\mbox{s}$. The inset figure in the right plot shows a scatter plot of the prompt and delayed NPE signals.

In addition, there is a 10~mm stainless-steel enclosure of the AmC source to attenuate the 60~keV gamma rays from $^{241}$Am decays. It result in neutron captures $^{56}$Fe,  $^{53}$Cr, and $^{58}$Ni, etc., releasing gamma rays with total energies from 7~MeV to 10~MeV. These high-energy gamma rays provide unique constraints to the energy scale determination.

\begin{figure}[htbp]
\centering
  \includegraphics[width=0.375\columnwidth]{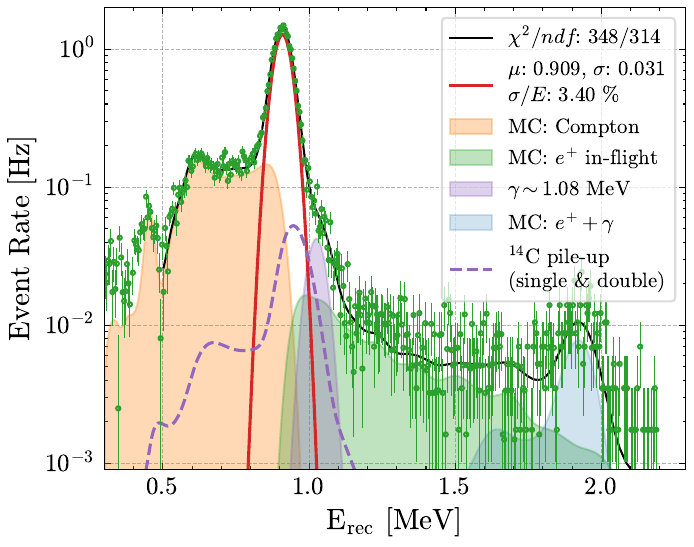}
  \includegraphics[width=0.55\columnwidth]{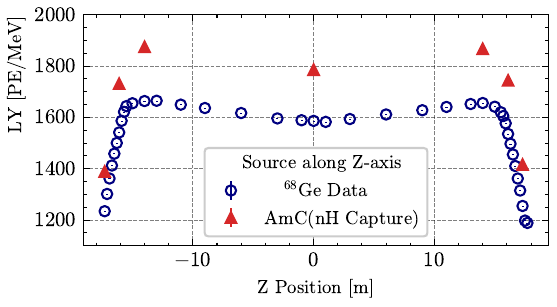}
  \caption{\label{fig:LY:data-mc}Left: reconstructed energy spectrum for a $^{68}\mbox{Ge}$ calibration source placed at the CD centre. Data points are compared to the simulation predictions.
  Right: LY measured with the $^{68}\mbox{Ge}$ and the AmC sources.}
\end{figure}

The left panel of Figure~\ref{fig:LY:data-mc} presents a typical reconstructed energy spectrum from the $^{68}\mbox{Ge}$ $\gamma$ calibration source. $^{68}\mbox{Ge}$ mainly undergoes EC to $^{68}\mbox{Ga}$ which decays $\beta^+$ decay to $^{68}\mbox{Zn}$, and the emitted positron promptly annihilates with an electron, producing two 511~keV $\gamma$s. The right panel of Figure~\ref{fig:LY:data-mc} shows the average LY of the JUNO LS in response to the total 1.022~MeV energy deposition from these $\gamma$s. The measured LY exceeds 1600~PE/MeV for source positions within $|z|<15~\mbox{m}$, surpassing initial expectations~\cite{JUNO:energy-reso:2025}. Additionally, as shown in Figure~\ref{fig:energy:AmC}, the LY for 2.223~MeV $\gamma$ is found to be 1785~PE/MeV at detector centre. The difference in LY values stems from the non-linear energy response of the LS, which will be discussed later.

The primary factor for the enhanced LY is the improved transparency of both the LS and water. The measured attenuation lengths are 20.6~m for the LS at 430~nm and greater than 70.0~m for water at 400~nm, outperforming the simulated values of 20.0~m and 40.0~m, respectively. After updating these key parameters in the simulation, the observed LY is well reproduced. It should be noted, however, that edge-related systematic effects are still under investigation, and further calibration data and analysis refinements are ongoing to improve the spatial response modeling.

\subsection{Vertex and energy reconstruction in the CD}

Several studies have been conducted on vertex and energy reconstruction using simulation data.
In~\cite{Wu:2018zwk,Liu:2018fpq,Li:2021oos,Huang:2021baf,Huang:2022zum}, a simultaneous vertex and energy reconstruction method named OMILREC was introduced, which constructs PMT time and charge maps using ACU and CLS calibration data.
Another approach, VTREP, following the concept presented in~~\cite{Takenaka:2025hgi}, utilizes $^{214}$Po background events in the BiPo cascade decay to build the charge map for energy reconstruction
Additionally, during the detector commissioning phase, a separate vertex reconstruction method called JVertex was developed.

Vertex reconstruction primarily relies on PMT timing information. After correcting for photon time-of-flight (TOF) and event start time, residual hit times of the PMT signals are obtained. The three vertex reconstruction methods have independent vertex searching algorithms:
\begin{itemize}
    \item JVertex employs a maximum-likelihood estimation method constructed from first hit times after time-of-flight subtraction. The event vertex is obtained by minimizing the negative log-likelihood, treating the trigger time as a free nuisance parameter. Light propagation effects are accounted for by introducing an effective refractive index equal to 1.60. The minimization uses all first hits recorded by the large PMTs and it is performed over corresponding time Probability Density Functions (PDFs). These PDFs are derived from AmC calibration data deployed with the ACU system and are constructed using only Dynode PMTs, which provide the best timing performance. No dependence on the calibration source position is observed. Separate PDFs are produced according to the detected photon multiplicity, to account for cases where multiple photons are registered by the same PMT within its time resolution.
    \item OMILREC also employed a time likelihood function which can be constructed using the observed first hit time of PMTs and the residual hit time PDF to reconstruct the vertex. The overall strategy is similar to that of JVertex. However, both Dynode and MCP PMTs were used, and their time PDFs were constructed from $^{68}$Ge source data at the CD centre separately. In addition, to improve the vertex reconstruction performance for the R$>$15~m region where total reflection could occur when scintillation photons enter water, the charge information of the PMTs was also used to constrain the vertex. A charge likelihood function was constructed using the observed and expected PMTs' charge, the latter of which highly depends on the event  vertex. The combined time and charge likelihood functions were maximized to obtain the most probable vertex.
     \item VTREP searches for the point at which the distribution of residual times becomes the sharpest. It starts with an initial vertex position estimated using the charge- and time-weighted barycentre of the fired PMT positions. Because photons that arrive later at the PMTs are more affected by scattering and internal reflections during propagation, PMTs with late hit times are assigned smaller weights in the initial (barycentre) vertex estimation. The final vertex estimate uses the summation over PMTs with residual times within $\pm$30~ns of the residual timing distribution to suppress contributions from PMT dark noise, as well as from scattered and reflected photons.

\end{itemize}

\begin{figure}[htbp]
\centering
  \includegraphics[width=0.95\columnwidth]{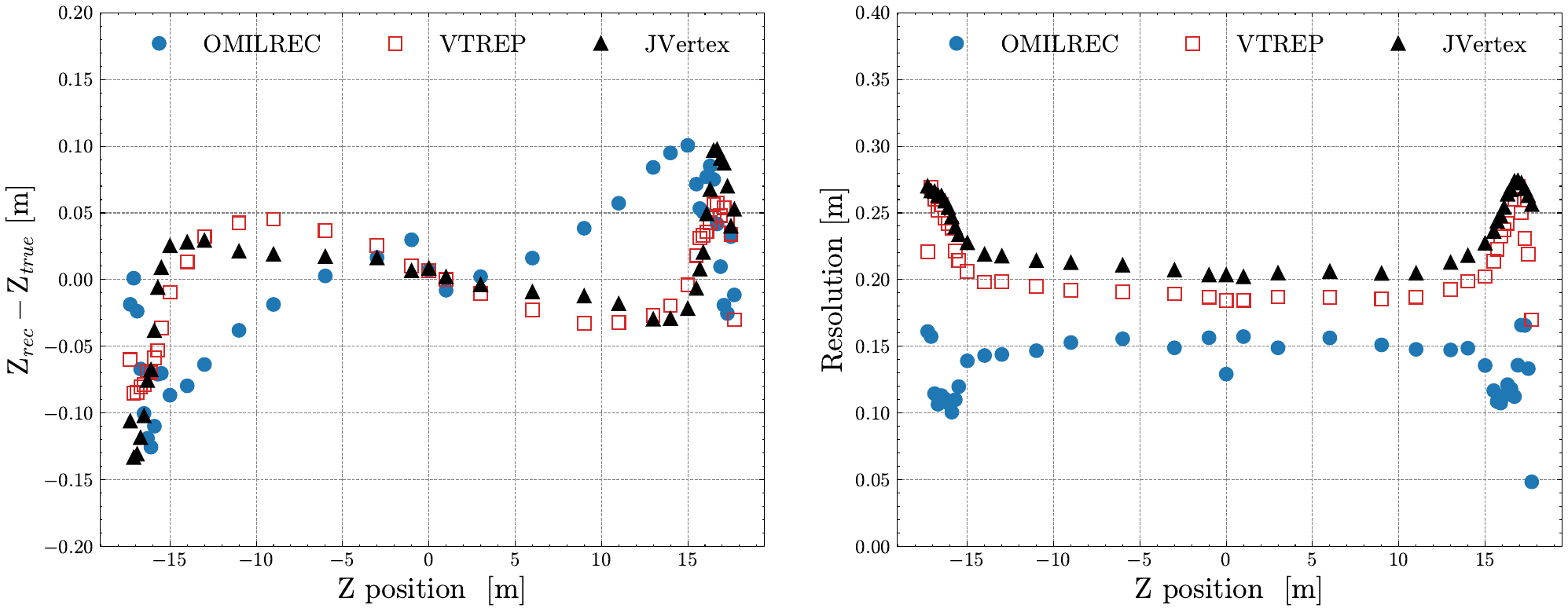}
  \caption{\label{fig:vertex:biasresolution} Bias and resolution of reconstructed $z$ for the $^{68}$Ge source deployed via ACU.  }
\end{figure}

The reconstruction performance was evaluated using a $^{68}$Ge calibration source deployed along the $z$-axis. Figure~\ref{fig:vertex:biasresolution} presents the bias and resolution of the reconstructed vertex at different $z$ positions. The results show that the vertex bias for all three methods remains within 10 cm across almost the entire detector. OMILREC demonstrates superior vertex resolution, particularly near the detector edge, which is attributed to its incorporation of PMT charge information.

The OMILREC and VTREP algorithms further reconstruct particle energy using the observed PMT charge information. The energy is obtained by searching for the value that maximizes the following likelihood function:
\begin{align}
L(E_{\rm rec}) &= \prod_{j}^{\rm Unhit~LPMTs}P_{j}({\rm unhit}|\mu_{j, \rm exp}) \times\prod_{i}^{\rm Hit~LPMTs}P_{i}(Q_{i, \rm obs}|\mu_{i, \rm exp}), \label{eq:energylikelihood} \\
\mu_{i,{\rm exp}} &= \mu_{i, {\rm S}}+\mu_{i, \rm dark} \nonumber \\
\end{align}
Here, $P({\rm unhit}|\mu)$ denotes the probability that no photons are detected for a given predicted charge $\mu$, while $P(Q|\mu)$ represents the probability of observing a charge $Q$ for the predicted charge $\mu$.
The predicted charge consists of contributions from scintillation and Cherenkov light $(\mu_{\rm S})$, as well as from dark noise hits $(\mu_{\rm dark})$.

Since the predicted light intensity from a particle depends on the spatial relationship between the particle and the PMT positions, it is tabulated as a map~(OMILREC) or function~(VTREP ) of the vertex position and the relative angle to the PMT.
OMILREC uses the $^{68}$Ge calibration source along the ACU and $\alpha$ events from $^{214}$Po distributed in the whole volume to construct the map, while VTREP uses $^{214}$Po events to obtain the function.
The probability density functions, $P({\rm unhit}|\mu)$ and $P(Q|\mu)$, represent the PMT charge response at different light intensity levels and are calibrated using laser runs with varying light intensities.

\begin{figure}[htbp]
\centering
  \includegraphics[width=0.95\columnwidth]{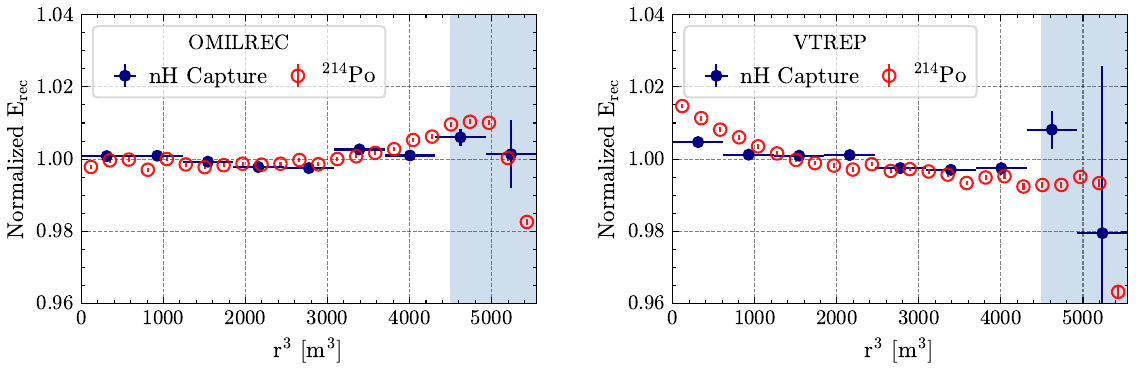}
  \caption{\label{fig:energy:nonuniformity} The residual spatial dependence of reconstructed energies for OMILREC~(left) and VTREP~(right). Reconstructed energy is uniform to better than $\pm$1\% in the $R < 16.5\ \mathrm{m}$ region. }
\end{figure}

The residual spatial dependence of the energy reconstruction, i.e., the energy non-uniformity, has been quantified using $^{214}$Po $\alpha$ decays and neutron captures from reactor antineutrino IBD reactions, as shown in Figure~\ref{fig:energy:nonuniformity}. Within the central region of $R < 16.5\ \mathrm{m}$, the reconstructed energies for both $\alpha$ particles and $\gamma$ exhibit a uniformity better than $\pm$1\% for both reconstruction methods.

\subsection{Energy non-linearity}

In LS detectors, the reconstructed energy does not depend linearly with the particle's deposited energy, due to particle-dependent quenching effects and the presence of Cherenkov radiation. Additional non-linearity can be introduced during the charge reconstruction of PMT waveforms~\cite{Huang:2017abb}. The nonlinear energy response has been extensively studied in experiments such as Daya Bay~\cite{DayaBay:2019fje}, Borexino~\cite{Borexino:2013zhu}, and Double Chooz~\cite{DoubleChooz:2019qbj}. Owing to the shared LAB-based liquid scintillator formulation, JUNO is expected to exhibit an intrinsic nonlinear response similar to that observed in Daya Bay. Furthermore, the charge reconstruction performance of the large PMTs is being verified by cross-checking with the small PMT system based on counting hit PMTs and thus linear in this energy regime due to their photon-counting operation.

\begin{figure}[htbp]
\centering
  \includegraphics[width=0.45\columnwidth]{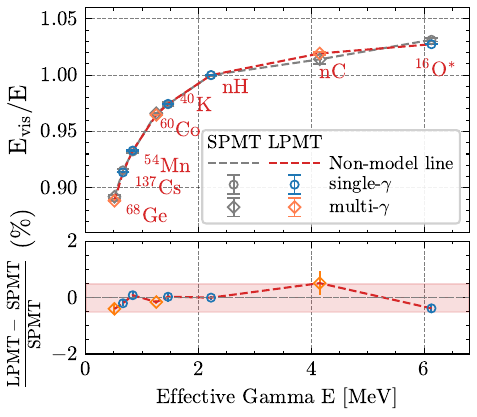}
  \includegraphics[width=0.45\columnwidth]{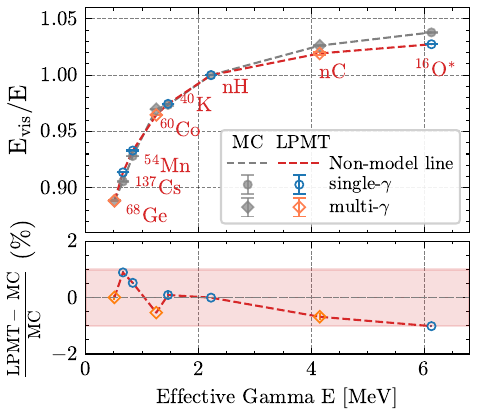}
  \caption{\label{fig:energy:nonlinearity} Left: non-linearity energy response measured by the large PMT and small PMT systems, indicating less than $\pm$0.5\% instrumental non-linearity in the reactor anti-neutrino energy region. Right: comparison of the $\gamma$ energy non-linearity between data and simulation, with a consistency better than $\pm$1\%.}
\end{figure}

The energy non-linearity, defined as the ratio of the visible energy to the deposited energy ($E_{\rm vis}/E$), was studied using various $\gamma$ calibration sources deployed at the detector centre. Here, $E_{\rm vis}$ is quantified as the total collected charge divided by a conversion factor anchored to the 2.223~MeV $\gamma$ from neutron capture on hydrogen, which is normalized to 1. $E_{\rm dep}$ represents the total energy deposited by the $\gamma$. The non-linearity is plotted as a function of the effective $\gamma$ energy, defined as the energy of a single $\gamma$ for mono-energetic sources, or the average energy per $\gamma$ for sources emitting multiple $\gamma$ per decay.
Figure~\ref{fig:energy:nonlinearity}(left) shows the energy non-linearity measured by large and small PMT systems. The consistency of the two measurements at the level of the current $\pm$0.5\% uncertainty indicates that any instrumental nonlinearity in the response of the large PMTs is below this level. This is confirmed by dedicated laser calibration runs where the large PMT response is directly benchmarked against the linear response of the small PMT system. Figure ~\ref{fig:energy:nonlinearity}(right) shows an excellent agreement within $\pm$1\% between data and simulation.

The Energy of a reactor antineutrino is primarily carried by the positron in IBD reactions. Kinetic energy deposit of positrons is  similar to electrons, while the annihilation $\gamma$'s deposit energy primarily via Compton scatterings with electrons. Thus, the non-linear energy response model for ($\gamma$, $e^-$, $e^+$) is built based on electrons. The total light yield, $L$, for a given deposited energy, $E_{\rm dep}$, is modelled as the sum of two components: the quenched scintillation light, $L_S$, and the Cherenkov light, $L_C$.

\begin{equation}
LY(E_{\rm dep}) =  LY_{\rm S}(E_{\rm dep}) + f_{\rm C} \cdot LY_{\rm C}(E_{\rm dep})~\mbox{,}
\end{equation}
where $f_C$ is the Cherenkov effective scaling factor.
The scintillation light quenching is modeled using Birks' law. The light yield per unit path length, $dLY_{\rm S}/dx$, is given by:

\begin{equation}
\frac{dLY_{\rm S}}{dx} = \frac{S \cdot dE_{\rm dep}/dx}{1 + k_{\rm B} \cdot dE_{\rm dep}/dx + k_{\rm C} \cdot (dE_{\rm dep}/dx)^2}~\mbox{,}
\end{equation}
where $S$ is the scintillation efficiency, $dE_{\rm dep}/dx$ is the stopping power of the particle in the scintillator, and $k_{\rm B}$ and $k_{\rm C}$ are Birks' constants. The total quenched scintillation light, $L_S$, is obtained by integrating this expression over the particle's track by summing over discrete energy depositions from the JUNO Monte Carlo simulations and/or ESTAR~\cite{berger1999ESTAR} database data. The Cherenkov light contribution, $LY_{\rm C}$, is obtained from JUNO Monte Carlo simulations.

\begin{figure}[htbp]
\centering
  \includegraphics[width=0.95\columnwidth]{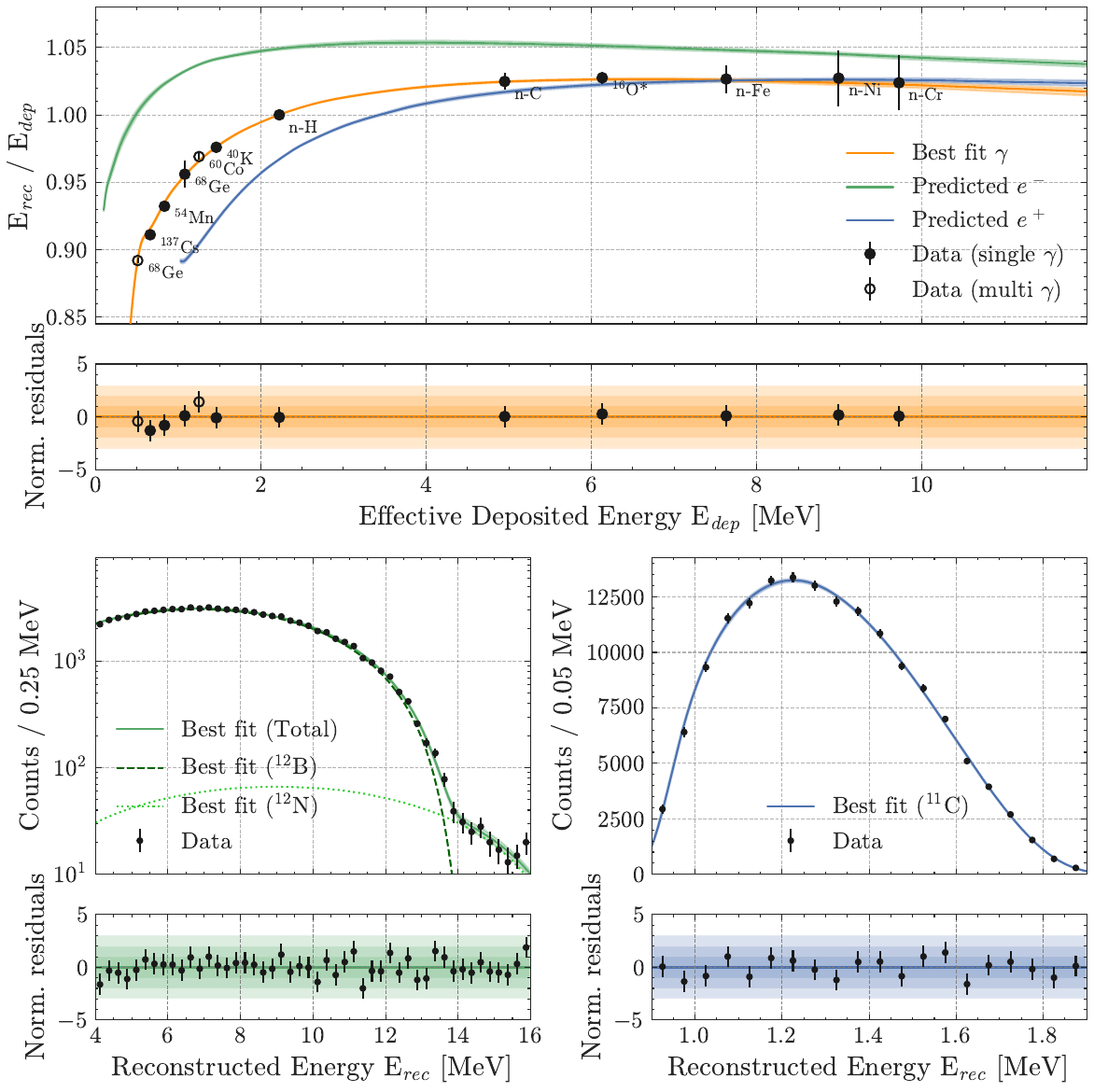}
  \caption{\label{fig:energy:NL-as} Energy non-linearity fitting results. Every study simultaneously fit the $\gamma$ calibration sources, cosmogenic $^{12}$B and $^{11}$C energy spectra. Top: $\gamma$ calibration sources non-linearity. Bottom left: $^{12}$B energy spectrum. Bottom right:
  the $^{11}$C energy spectrum.}
\end{figure}

Once the total light $LY$ is computed for any particle, the non-linearity is defined in eq.~(\ref{eq:nl}) as the ratio of the light yield to the deposited energy. To link the energy scales, all particle-specific non-linearity curves ($\gamma$, $e^-$, $e^+$) are normalized to a single anchor point. In the top plot of Figure~\ref{fig:energy:NL-as}, all scales are normalized to the LY of the $\gamma$ from n-H capture ($E_{\rm dep} = 2.223$ MeV). In addition, the instrumental non-linearity $f_{\rm inst.}$ is introduced and modelled with a simple linear term $f_{\rm inst.}(E_{\rm vis}) = 1 - k_I \cdot E_{\rm vis}$, where the parameter $k_I$ is determined from data and constrained with the small PMT system. The normalized non-linearity for any particle becomes:

\begin{equation}
nl(E_{\rm dep})=\frac{E_{\rm rec}}{E_{\rm dep}} = \frac{LY(E_{\rm dep})}{E_{\rm dep}}\cdot \frac{E_{\text{anchor}}}{L_{\text{anchor}}}\cdot f_{\rm inst.}~\mbox{,}
\label{eq:nl}
\end{equation}
where $E_{\text{anchor}} = 2.223~\mbox{MeV}$ and $LY_{\text{anchor}}$ is the light yield for that specific particle and energy.

The data used in the determination of energy non-linearity consist of multiple $\gamma$ calibration sources in the detector centre, the $\beta^-$ spectrum from cosmogenic $^{12}$B decays in a $R<$16.5~m fiducial volume, and the $\beta^+$ spectrum from cosmogenic $^{11}$C decays in a $R<$16.5~m fiducial volume.
The top plot of Figure~\ref{fig:energy:NL-as} shows the reconstructed
energy for single and multiple $\gamma$ calibration sources divided by
the expected deposited energy.
The orange curve is a fit to the data modelling the $\gamma$ non-linearity predicted by the Monte Carlo, including the Cherenkov and quenching contributions as follows. The models use a pre-computed look-up table, derived from Geant4 simulations, that gives the Cherenkov yield $C(E_k)$ as a function of an electron's kinetic energy $E_k$. The model predictions also agree well with $^{11}$C and $^{12}$B spectra.

\subsection{Energy resolution}

\begin{figure}[htbp]
\centering
  \includegraphics[width=0.6\columnwidth]{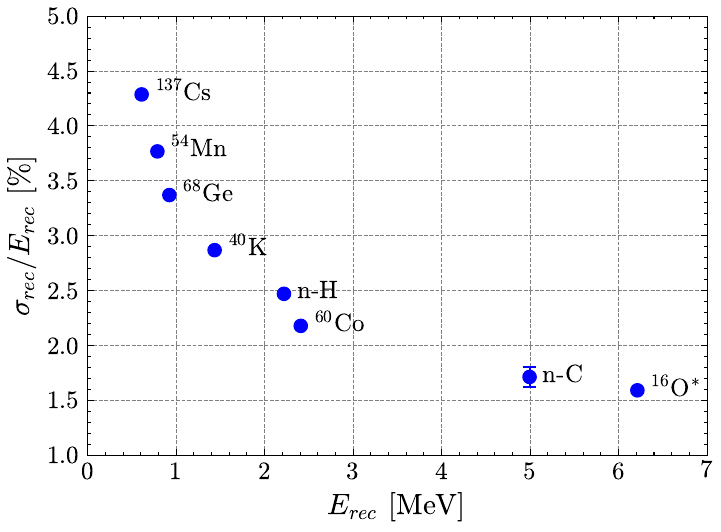}
  \caption{\label{fig:energy:gammaresolution} Energy resolution as a function of energy measured with $\gamma$ sources placed at the CD centre. }
\end{figure}

The energy resolution is a key parameter for determining the neutrino mass ordering. Figure~\ref{fig:energy:gammaresolution} shows the energy resolution achieved for $\gamma$ at the detector centre. The resolution for the $^{68}$Ge source is approximately 3.4\% (left plot of Figure~\ref{fig:LY:data-mc}).
If fitting the resolutions using the equation below:
\begin{equation}
\sigma_E/E = \sqrt{a^2/E+b^2},
\end{equation}
the stochastic term $a$ is $\sim$3.3\% and the constant term $b$ is $\sim$1\%.
These values may evolve because of on-going studies on the precise fitting of gamma spectra and the uncertainty estimate.
In addition, the energy resolution for $\gamma$ is generally degraded compared to that for electrons, due to the additional energy smearing introduced by Compton scattering processes.

%

The contributions of major factors to the total energy resolution budget were also evaluated, including the quenching effect, Cherenkov light contribution in scintillation processes, dark noise, and single photoelectron charge smearing, among others. Preliminary analyses suggest that the slightly worse energy resolution in data may be primarily attributed to partial treatment of $^{14}$C pileup in the $^{68}$Ge data, resulting from the intrinsic $^{14}$C concentration of (3 to 5)$\times 10^{-17}$~g/g in the liquid scintillator. Dedicated efforts are currently underway to improve the energy resolution.

\begin{figure}[htbp]
\centering
  \includegraphics[width=0.85\columnwidth]{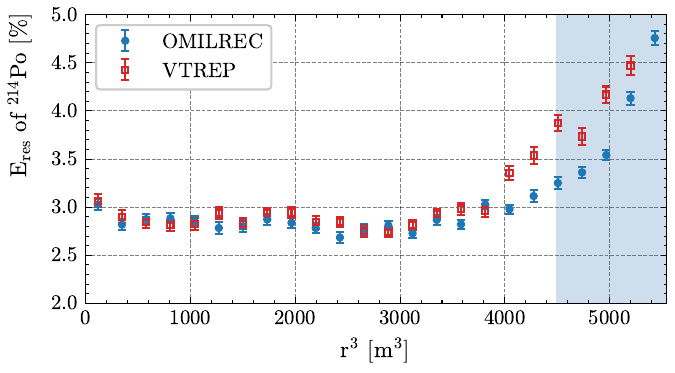}
  \caption{\label{fig:energy:alpharesolution} Energy resolution versus R$^3$ evaluated using $^{214}$Po with a reconstructed energy of about 0.93~MeV. }
\end{figure}

Additionally, the energy resolution across the entire detector volume has been systematically evaluated using $^{214}$Po events from the $^{238}$U decay chain. Owing to the pronounced quenching of $\alpha$ particles in the LS, the reconstructed energy of $^{214}$Po is approximately 0.93 MeV. The energy resolution is presented as a function of the radial volume ($R^3$) in Figure~\ref{fig:energy:alpharesolution}. Except for a slight degradation near the detector boundary, the resolution for $^{214}$Po reaches 2.8\%. A comparative analysis between these $\alpha$ events distributed in the whole volume and the $\gamma$ events at the centre is expected to refine our understanding of the energy resolution in the near future.

\section{Muon detection in JUNO}
Cosmic-ray muons traversing the liquid scintillator can produce spallation products such as $^9$Li, $^8$He, $^{12}$B, $^8$Li, $^8$B, and $^{11}$C. These unstable isotopes can mimic the correlated signals of reactor neutrino IBD events or the single-electron scattering signals induced by solar neutrinos.
Moreover, untagged muons in the WCD can generate fast-neutron backgrounds in the CD.

The WCD employs a regional trigger logic for the PMTs:
the detector sphere is divided into ten trigger zones, five in the upper and five in the lower hemisphere, arranged in petal-shaped sectors. In addition, 3 zones are defined for the three rings of PMTs mounted on the water pool walls.  A local trigger threshold is defined for each zone, requiring about 30 PMT hits to generate a trigger validation signal. In addition, a cross-zone trigger condition is applied to pairs of adjacent zones, where a combined total of approximately 40 fired PMTs is required. The specific thresholds are fine-tuned individually according to the noise characteristics of each zone. This hierarchical trigger design effectively reduces the overall detector threshold and improves the efficiency for muon detection.

To ensure high detection efficiency in the WCD, water transparency is of paramount importance. A dedicated device was developed for online monitoring of water transparency. Installed at the bottom of the water pool, the system is equipped with five 20-inch MCP-PMTs, LEDs, and optical fibers~\cite{Wang:2023iul}. After the pool was filled, the water transparency was continuously monitored using this setup.
\begin{figure}
  \centering
  \includegraphics[width=0.85\textwidth]{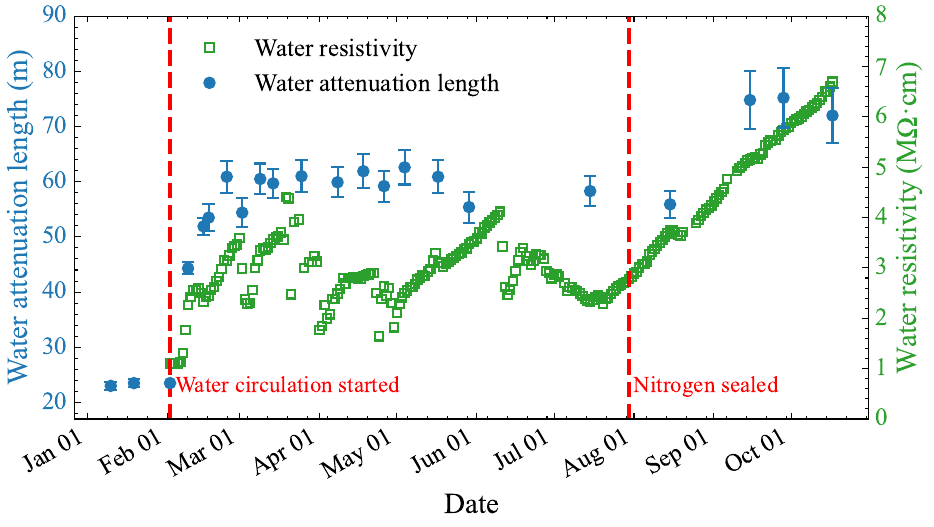}
  \caption{Time evolution of the water attenuation length in the WCD (blue circles). The first increase, at the end of February, occurred after water circulation began; the second, in early September, coincided with nitrogen sealing of the water pool. The corresponding water resistivity is superimposed for correlation (green square).}
  \label{fig:WCD:WAL}
\end{figure}
During the water filling stage (from 18~December~2024 to 1~February~2025), water circulation inside the detector was not yet active, resulting in a water attenuation length of approximately 25~m at a LED wavelength of 400~nm (see blue points in Figure \ref{fig:WCD:WAL}). Once water circulation began, the attenuation length increased sharply and eventually stabilized at around 60~m. A second rise in the attenuation length was observed at the start of the science runs, corresponding to the completion of full nitrogen sealing of the water pool. These improvements are consistent with the observed increase in water resistivity (overlaid in green in the plot of Figure~\ref{fig:WCD:WAL}).

\begin{figure}[htbp]
\centering
  \includegraphics[width=0.95\columnwidth]{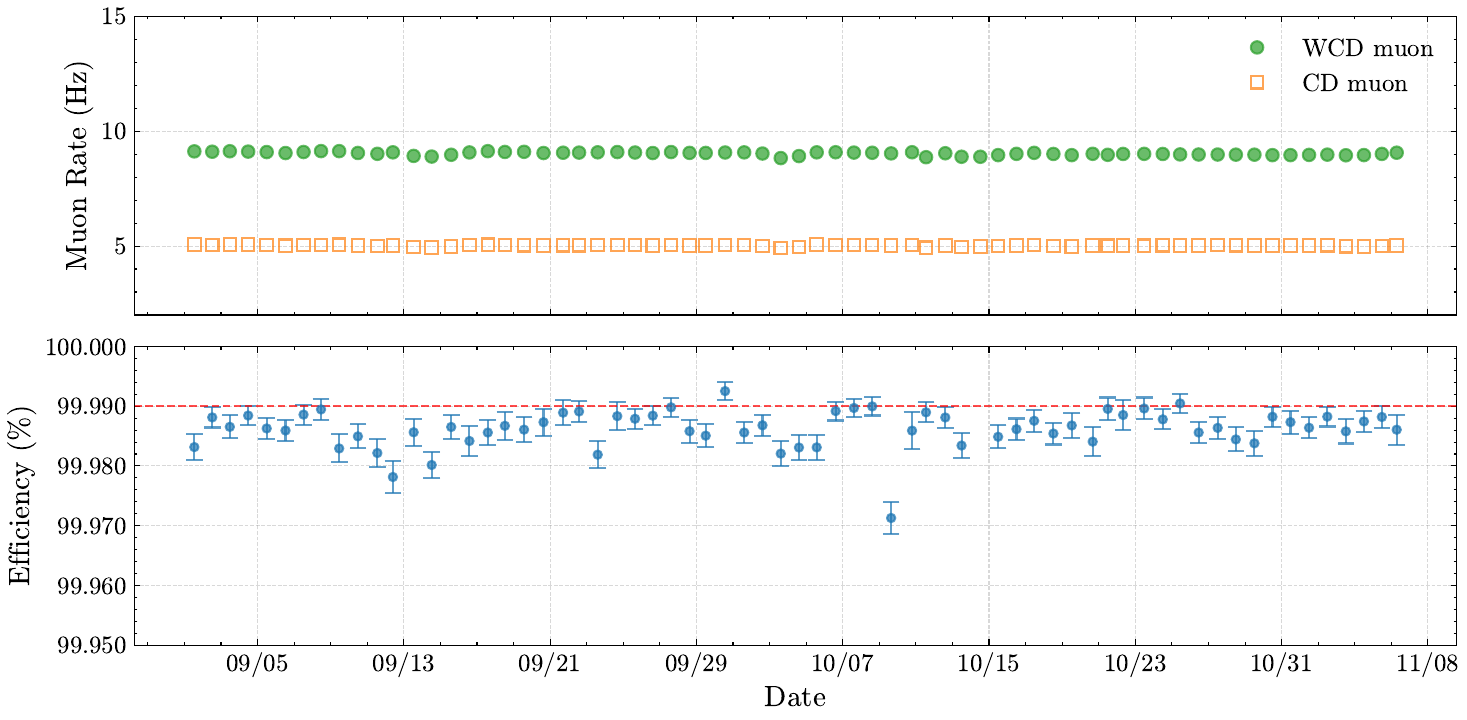}
    \caption{\label{fig:muon:tag-eff}
    Top panel: cosmic-ray muon rates in CD and WCD in September and October~2025. Bottom panel: Time evolution of WCD muon tagging efficiency for muons passing CD.}
\end{figure}

The top plot of Figure~\ref{fig:muon:tag-eff} shows the rate of muons tagged in the CD  ($\sim$5~Hz) and in the WCD ($\sim$9~Hz). The lower panel of the same figure shows the muon tagging efficiency in the WCD, calculated by studying the temporal correlation between events with a muon tagged in the WCD and those detected in the CD.
From 1 September to 6 November 2025, the WCD achieved a muon tagging efficiency nearly equal to 1, with a small geometrical inefficiency arising from the top chimney that links the CD to the calibration house.

\subsection{Muon track reconstruction}

According to track multiplicities, muons can be classified as single muons or muon bundles.
These muon bundles are often produced in the same hadronic interaction occurring a few tens of km away from the detector, and therefore, they tend to reach the JUNO detector with track directions almost parallel to each other.

In JUNO, the current muon classification method is a cluster-finding algorithm applied to the PMT charge. The basic principle is that a muon passing through the detector will create localized clusters of high charge on the PMTs, typically at the detector entry and exit points.
The algorithm identifies hit-clusters by searching for PMTs with high collected charge.

Several muon reconstruction methods are employed in JUNO.
The baseline method relies on the identification of local charge clusters and, for events with two or more clusters, reconstructs the muon track(s) assuming down-going muons and connecting the measured entry and exit points, evaluated as the charge-weighted positions of the PMTs within the identified clusters, with a straight lines.
\begin{figure}[htbp]
\centering
  \includegraphics[width=0.95\columnwidth]{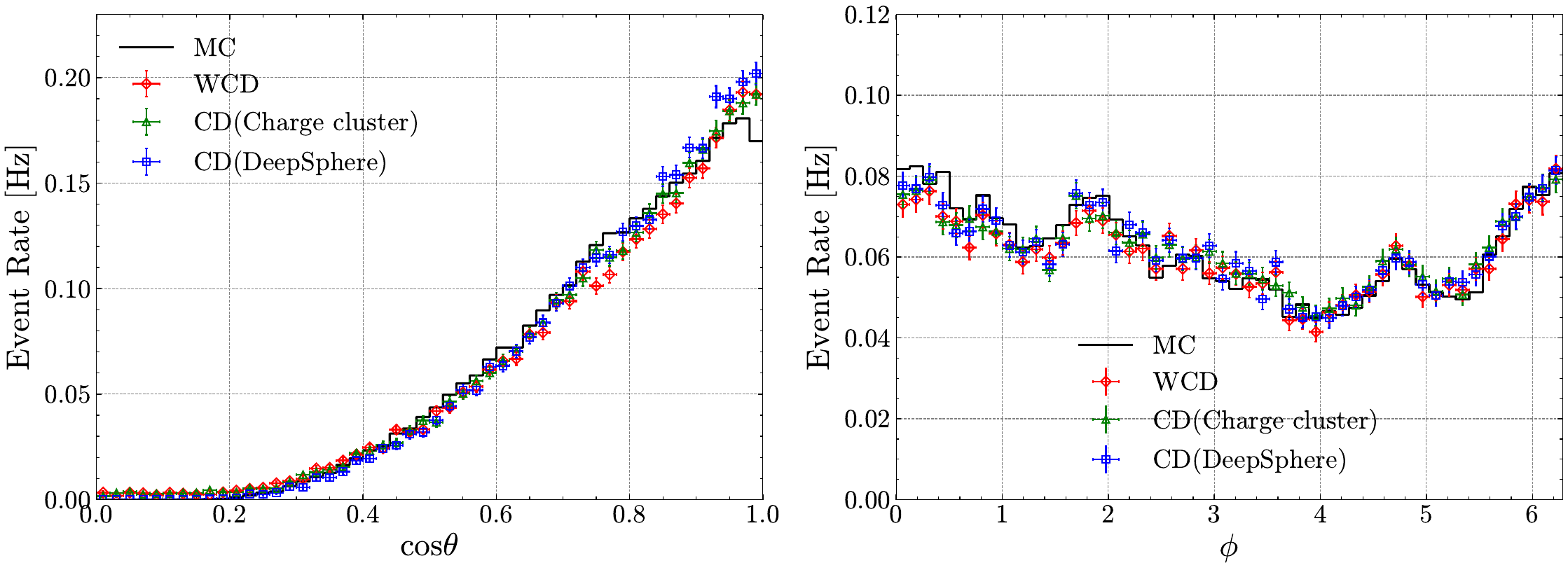}
    \caption{\label{fig:muon:angular}
    Left: Single muon $\cos{\theta}$ distribution.
    Right: Single muon $\phi$ angle distribution. Good agreement between different reconstruction algorithms and simulation is visible.}
\end{figure}
In addition, a few Machine Learning based methods were developed for reconstructing muons in atmospheric neutrino interactions in JUNO~\cite{Yang:2023rbg,Liu:2025fry}.
Two models were employed in this study: one based on DeepSphere~\cite{Perraudin:2018rbt}, a graph convolutional neural network dedicated to processing sperical image-like data directly, and a second algorithm based on CoAtNet~\cite{dai2021coatnet}, a hybrid architecture that integrates convolutional operations and self-attention mechanisms.
The DeepSphere-based model groups PMT charge signals into pixels following the HEALPix sampling scheme~\cite{HEALPix} before feeding them into the network.
The CoAtNet-based model maps the PMT charge information onto a two-dimensional $(\theta_{\mbox{PMT}},\phi_{\mbox{PMT}})$ grid, where $\theta_{\mbox{PMT}}$ and $\phi_{\mbox{PMT}}$ denote the zenith and azimuthal angles of each PMT in the detector, respectively.
Both models are trained to identify the entry and exit points of the muons, and the line connecting the two are defined as the reconstructed muon track.

Figure~\ref{fig:muon:angular} shows the angular distributions in terms of the zenith (left) and azimuth (right) angles of the reconstructed single muons. The plot shows the WCD only reconstructed muons using the baseline method, and the muons reconstructed in the CD with two different methods.
To simulate the muon flux and angular distributions in the underground experimental hall, high- and low-precision topographic maps of the Dashi Hill region were combined to generate a realistic, digitized mountain profile. Sea-level muons were simulated using a modified Gaisser formula~\cite{Gaisser:2016uoy,Tang:2005zz}, which provides a more accurate description of the low-energy spectrum and large-zenith-angle muons. The rock traversal length for each muon was computed via interpolation based on the digitized terrain. The muons were then propagated through the rock using the MUSIC code~\cite{MUSIC1,MUSIC2} to obtain the underground Monte Carlo muon sample. As clearly shown in Figure~\ref{fig:muon:angular}, the simulation exhibits good agreement with the experimental data, indicating that the mountain profile is accurately represented in the simulation.

\begin{figure}[htbp]
\centering
  \includegraphics[width=0.925\columnwidth]{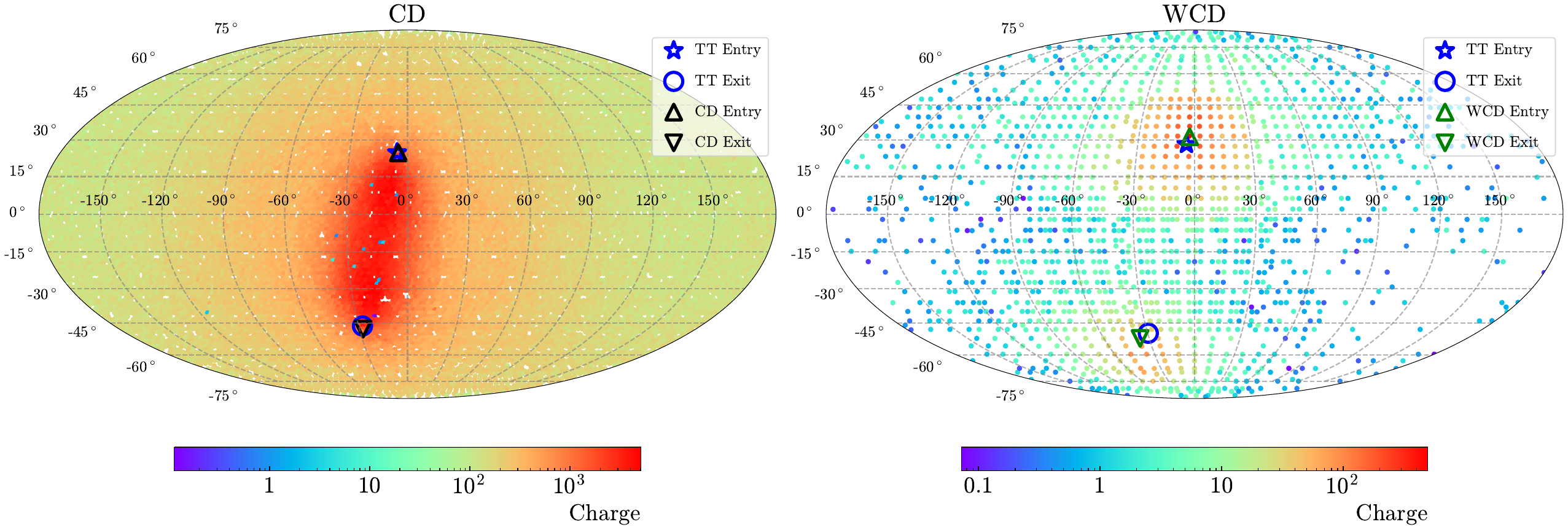}
    \caption{\label{fig:muon:display}
    JUNO through-going muon event display.
    The same event is reconstructed in the CD (left plot) and in the WCD (right plot). The colour code shows the amount of charge collected in the PMTs.}
\end{figure}
Finally, Figure~\ref{fig:muon:display} illustrates an event display of a through-going muon. The entry and exit points reconstructed in the CD (left) and WCD (right) are compared with the predictions from the TT. The muon track reconstruction is expected to further improve as additional TT data become available in the near future.

\section{Conclusion}

JUNO, with its 20 kton of liquid scintillator target and 35 kton of water Cherenkov veto detector is the largest liquid scintillator based experiment ever built.
This paper presented a description of the final configuration of the experiment and its physics performances measured  first in the two months pure water phase, afterwards during the six months liquid scintillator filling phase, and finally during the first two months of science runs.

The majority of detector subsystems have entered operation meeting or exceeding design expectations. In particular, the photon detection system achieves state-of-the-art performances for a large target-mass neutrino detector: the 20-inch PMTs exhibit a low {\marka dead-channel fraction of 0.11\%} and an {\marka average dark noise rate below 23 kHz}.

All auxiliary JUNO facilities, the high purity water plant and the LS purification systems have performed according to specifications allowing to start operation in a smooth way. The calibration system is being used regularly to perform the timing and energy calibration of the detector, to determine the energy scale and to correct for non-linearities and non-uniformities in the overall detector response.

A careful selection of the materials, a cleaning program for the detector surfaces and operation of the high purity water plant and the LS purification systems have allowed for remarkable results in the radiopurity of the LS target. The $^{238}\mbox{U}$ and $^{232}\mbox{Th}$ concentrations have been measured as  {\marka$(7.5\pm 0.9)\times 10^{-17}\mbox{g}/\mbox{g}$} and {\marka$(8.2\pm 0.7)\times 10^{-17}\mbox{g}/\mbox{g}$}, respectively, one order of magnitude better than the design requirements for NMO~\cite{JUNO:NMOsensitivity:2025} determination and fully compliant for its solar neutrino measurement program~\cite{JUNO:solar:2023}.
Moreover, the amount of $^{210}\mbox{Po}$ was measured at the end of the LS filling at about {\marka $5\times 10^4~\mbox{cpd}/\mbox{kton}$}, decreasing to an average {\marka $(4.3\pm0.3)\times 10^4~\mbox{cpd}/\mbox{kton}$} over the first two months of science runs.

The energy response of the JUNO detector is well understood. The non-linearity has been calibrated with a precision better than 1\%, which is a remarkable result achieved after only two months of data taking.
In addition, a high light yield of 1785~PE/MeV has been achieved at the centre of the central detector, exceeding the expectations.
We have reached an {\marka energy resolution of 3.4\%} for the two 0.511~MeV $\gamma$ from $^{68}$Ge, and ongoing efforts, including $^{14}$C pileup discrimination, development of advanced reconstruction methods, and a refined understanding of Cherenkov contributions, are expected to improve it in the near future.
We believe that the JUNO detector has met its design objectives and is now fully ready to achieve its physics goals.

\section*{Acknowledgments}
We gratefully acknowledge the continued cooperation and support of the China General Nuclear Power Group in the construction and operation of the JUNO experiment.

We acknowledge financial and institutional support from the Chinese Academy of Sciences, the National Key R\&D Program of China, the People's Government of Guangdong Province, and the Tsung--Dao Lee Institute of Shanghai Jiao Tong University in China.

We appreciate the contributions from the Institut National de Physique Nucl\'eaire et de Physique des Particules (IN2P3) in France, the Istituto Nazionale di Fisica Nucleare (INFN) in Italy, the Fonds de la Recherche Scientifique (F.R.S.--FNRS) and the Institut Interuniversitaire des Sciences Nucl\'eaires (IISN) in Belgium, and the Conselho Nacional de Desenvolvimento Cient\'ifico e Tecnol\'ogico (CNPq) in Brazil.

We also acknowledge the support of the Agencia Nacional de Investigaci\'on y Desarrollo (ANID) and the ANID--Millennium Science Initiative Program (ICN2019044) in Chile; the European Structural and Investment Funds, the Ministry of Education, Youth and Sports, and the Charles University Research Center in the Czech Republic; the Deutsche Forschungsgemeinschaft (DFG), the Helmholtz Association, and the Cluster of Excellence PRISMA+ in Germany; and the Joint Institute for Nuclear Research (JINR) and Lomonosov Moscow State University in Russia.

We further thank the Slovak Research and Development Agency in the Slovak Republic, the National Science and Technology Council (NSCT) and MOE in Taiwan, China, the Program Management Unit for Human Resources \& Institutional Development, Research and Innovation (PMU--B), Chulalongkorn University, and Suranaree University of Technology in Thailand, the Science and Technology Facilities Council (STFC) in the United Kingdom, and the University of California at Irvine and the National Science Foundation (NSF) in the United States.

We also acknowledge the computing resources provided by the Chinese Academy of Sciences, IN2P3, INFN, and JINR, which are essential for data processing and analysis within the JUNO Collaboration.

\bibliographystyle{unsrt}
\bibliography{sample}

\end{document}

%% file: author.tex
\author[7]{Angel Abusleme}
\author[52]{Thomas Adam}
\author[58]{Kai Adamowicz}
\author[12]{David Adey}
\author[74]{Shakeel Ahmad}
\author[74]{Rizwan Ahmed}
\author[49]{Timo Ahola}
\author[64]{Sebastiano Aiello}
\author[23]{Fengpeng An}
\author[12]{Guangpeng An}

% ...
\affil[7]{Pontificia Universidad Cat\'{o}lica de Chile, Santiago, Chile}
% ...
\affil[12]{Institute of High Energy Physics, Beijing, China}
% ...
\affil[23]{Sun Yat-sen University, Guangzhou, China}
% ...
\affil[49]{University of Jyvaskyla, Department of Physics, Jyvaskyla, Finland}
% ...
\affil[52]{IPHC, Universit\'{e} de Strasbourg, CNRS/IN2P3, F-67037 Strasbourg, France}
% ...
\affil[58]{III. Physikalisches Institut B, RWTH Aachen University, Aachen, Germany}
% ...
\affil[64]{INFN Catania and Dipartimento di Fisica e Astronomia dell Universit\`{a} di Catania, Catania, Italy}
% ...
\affil[74]{Pakistan Institute of Nuclear Science and Technology, Islamabad, Pakistan}
% ...


%% file: main.bbl
\begin{thebibliography}{10}

\bibitem{Zhan:2008id}
Liang Zhan, Yifang Wang, Jun Cao, and Liangjian Wen.
\newblock {Determination of the Neutrino Mass Hierarchy at an Intermediate
  Baseline}.
\newblock {\em Phys. Rev. D}, 78:111103, 2008.

\bibitem{Zhan:2009rs}
Liang Zhan, Yifang Wang, Jun Cao, and Liangjian Wen.
\newblock {Experimental Requirements to Determine the Neutrino Mass Hierarchy
  Using Reactor Neutrinos}.
\newblock {\em Phys. Rev. D}, 79:073007, 2009.

\bibitem{JUNO:2015zny}
Fengpeng An et~al.
\newblock {Neutrino Physics with JUNO}.
\newblock {\em J. Phys. G}, 43(3):030401, 2016.

\bibitem{JUNO:NMOsensitivity:2025}
Angel Abusleme et~al.
\newblock {Potential to identify neutrino mass ordering with reactor
  antineutrinos at JUNO}.
\newblock {\em Chin. Phys. C}, 49(3):033104, 2025.

\bibitem{JUNO:NuOscPar:2022}
Angel Abusleme et~al.
\newblock {Sub-percent precision measurement of neutrino oscillation parameters
  with JUNO}.
\newblock {\em Chin. Phys. C}, 46(12):123001, 2022.

\bibitem{JUNO:DetPhys:2022}
Angel Abusleme et~al.
\newblock {JUNO physics and detector}.
\newblock {\em Prog. Part. Nucl. Phys.}, 123:103927, 2022.

\bibitem{JUNO:CCSN:2024}
Angel Abusleme et~al.
\newblock {Real-time monitoring for the next core-collapse supernova in JUNO}.
\newblock {\em JCAP}, 01:057, 2024.

\bibitem{JUNO:DSNB:2022}
Angel Abusleme et~al.
\newblock {Prospects for detecting the diffuse supernova neutrino background
  with JUNO}.
\newblock {\em JCAP}, 10:033, 2022.

\bibitem{JUNO:geoneutrino:2025}
Thomas Adam et~al.
\newblock {Prospects for geoneutrino detection in JUNO}.
\newblock 2025.

\bibitem{JUNO:nuatmLowEnergy:2021}
Angel Abusleme et~al.
\newblock {JUNO sensitivity to low energy atmospheric neutrino spectra}.
\newblock {\em Eur. Phys. J. C}, 81:10, 2021.

\bibitem{JUNO:solar8B:2021}
Angel Abusleme et~al.
\newblock {Feasibility and physics potential of detecting $^8$B solar neutrinos
  at JUNO}.
\newblock {\em Chin. Phys. C}, 45(2):023004, 2021.

\bibitem{JUNO:solar:2023}
Angel Abusleme et~al.
\newblock {JUNO sensitivity to $^{7}$Be, pep, and CNO solar neutrinos}.
\newblock {\em JCAP}, 10:022, 2023.

\bibitem{JUNO:2022jkf}
Jie Zhao et~al.
\newblock {Model-independent Approach of the JUNO $^{8}$B Solar Neutrino
  Program}.
\newblock {\em Astrophys. J.}, 965(2):122, 2024.

\bibitem{JUNO:pdecay:2023}
Angel Abusleme et~al.
\newblock {JUNO Sensitivity on Proton Decay $p\to \bar\nu K^+$ Searches}.
\newblock {\em Chin. Phys. C}, 47(11):113002, 2023.

\bibitem{JUNO:DarkMatterGalacticHalo:2023}
Angel Abusleme et~al.
\newblock {JUNO sensitivity to the annihilation of MeV dark matter in the
  galactic halo}.
\newblock {\em JCAP}, 09:001, 2023.

\bibitem{JUNO:PMTmassTesting:2022}
Angel Abusleme et~al.
\newblock {Mass testing and characterization of 20-inch PMTs for JUNO}.
\newblock {\em Eur. Phys. J. C}, 82(12):1168, 2022.

\bibitem{JUNO:CalibStrategy:2021}
Angel Abusleme et~al.
\newblock {Calibration Strategy of the JUNO Experiment}.
\newblock {\em JHEP}, 03:004, 2021.

\bibitem{Cao_2021}
Chuanya~Cao et~al.
\newblock Mass production and characterization of 3-inch pmts for the juno
  experiment.
\newblock {\em Nucl. Instrum. Meth. A}, 1005:165347, 2021.

\bibitem{JUNO:sPMTinstrumentation:2025}
Jilei Xu et~al.
\newblock {Instrumentation of JUNO 3-inch PMTs}.
\newblock {\em arXiv}, 10 2025.

\bibitem{DayaBay:detector-paper:2015}
F.~P. An et~al.
\newblock {The Detector System of The Daya Bay Reactor Neutrino Experiment}.
\newblock {\em Nucl. Instrum. Meth. A}, 811:133--161, 2016.

\bibitem{Zhang:2021ikn}
G.~Zhang et~al.
\newblock {Addendum: The study of active geomagnetic shielding coils system for
  JUNO}.
\newblock {\em JINST}, 16(12):A12001, 2021.

\bibitem{JUNO:TTnim:2023}
Angel Abusleme et~al.
\newblock {The JUNO experiment Top Tracker}.
\newblock {\em Nucl. Instrum. Meth. A}, 1057:168680, 2023.

\bibitem{OPERA:TT}
T.~Adam et~al.
\newblock {The OPERA experiment target tracker}.
\newblock {\em Nucl. Instrum. Meth. A}, 577:523--539, 2007.

\bibitem{JUNO:upwater:2024}
T.~Y. Guan et~al.
\newblock {Development of low-radon ultra-pure water for the Jiangmen
  Underground Neutrino Observatory}.
\newblock {\em Nucl. Instrum. Meth. A}, 1063:169244, 2024.

\bibitem{JUNO:2021kxb}
Angel Abusleme et~al.
\newblock {Radioactivity control strategy for the JUNO detector}.
\newblock {\em JHEP}, 11:102, 2021.

\bibitem{Li:2024lcy}
C.~Li, B.~Wang, Y.~Liu, C.~Guo, Y.~P. Zhang, J.~C. Liu, Q.~Tang, T.~Y. Guan,
  and C.~G. Yang.
\newblock {Developing a {\ensuremath{\mu}}Bq/m3 level 226Ra concentration in
  water measurement system for the Jiangmen Underground Neutrino Observatory}.
\newblock {\em Nucl. Instrum. Meth. A}, 1063:169257, 2024.

\bibitem{JUNO-DYB:LScomposition:2021}
A.~Abusleme et~al.
\newblock {Optimization of the JUNO liquid scintillator composition using a
  Daya Bay antineutrino detector}.
\newblock {\em Nucl. Instrum. Meth. A}, 988:164823, 2021.

\bibitem{Zhu:OpticalPurifiction:2022}
Z.~Zhu et~al.
\newblock {Optical purification pilot plant for JUNO liquid scintillator}.
\newblock {\em Nucl. Instrum. Meth. A}, 1048:167890, 2023.

\bibitem{Landini:distillation:2024}
C.~Landini et~al.
\newblock {Distillation and gas stripping purification plants for the JUNO
  liquid scintillator}.
\newblock {\em Nucl. Instrum. Meth. A}, 1069:169887, 2024.

\bibitem{Ye:WaterExtraction:2021}
Jiaxuan Ye et~al.
\newblock {Development of water extraction system for liquid scintillator
  purification of JUNO}.
\newblock {\em Nucl. Instrum. Meth. A}, 1027:166251, 2022.

\bibitem{Ling:NitrogenPlant:2024}
Xin Ling et~al.
\newblock {JUNO high purity nitrogen plant}.
\newblock {\em Appl. Radiat. Isot.}, 208:111305, 2024.

\bibitem{JUNO:2021wzm}
Angel Abusleme et~al.
\newblock {The design and sensitivity of JUNO{\textquoteright}s scintillator
  radiopurity pre-detector OSIRIS}.
\newblock {\em Eur. Phys. J. C}, 81(11):973, 2021.

\bibitem{osiris:ana:2025}
Angel Abusleme et~al.
\newblock {Analysis of scintillator background levels with JUNO's pre-detector
  OSIRIS}.
\newblock {\em paper in preparation}.

\bibitem{calib:ACU}
Jiaqi Hui et~al.
\newblock {The automatic calibration unit in JUNO}.
\newblock {\em JINST}, 16(08):T08008, 2021.

\bibitem{calib:GT}
Yuhang Guo et~al.
\newblock {Design of the Guide Tube Calibration System for the JUNO
  experiment}.
\newblock {\em JINST}, 14(09):T09005, 2019.

\bibitem{calib:CLS}
Yuanyuan Zhang et~al.
\newblock {Cable loop calibration system for Jiangmen Underground Neutrino
  Observatory}.
\newblock {\em Nucl. Instrum. Meth. A}, 988:164867, 2021.

\bibitem{JUNO:USS1}
Guo-Lei Zhu et~al.
\newblock {Ultrasonic positioning system for the calibration of central
  detector}.
\newblock {\em Nucl. Sci. Tech.}, 30(1):5, 2019.

\bibitem{JUNO:USS2}
Duo Teng et~al.
\newblock {Low-radioactivity ultrasonic hydrophone used in positioning system
  for Jiangmen Underground Neutrino Observatory}.
\newblock {\em Nucl. Sci. Tech.}, 33(6):76, 2022.

\bibitem{JUNO:ROV}
K.~Feng et~al.
\newblock {A novel remotely operated vehicle as the calibration system in
  JUNO}.
\newblock {\em JINST}, 13(12):T12001, 2018.

\bibitem{JUNO:RadContrStrategy:2021}
Angel Abusleme et~al.
\newblock {Radioactivity control strategy for the JUNO detector}.
\newblock {\em JHEP}, 11:102, 2021.

\bibitem{Zhang:LSrefractiveindex:2024}
H.~S. Zhang et~al.
\newblock {Refractive index in the JUNO liquid scintillator}.
\newblock {\em Nucl. Instrum. Meth. A}, 1068:169730, 2024.

\bibitem{Beretta:LSfluorescence:2025}
M.~Beretta et~al.
\newblock {Fluorescence emission of the JUNO liquid scintillator}.
\newblock {\em JINST}, 20(05):P05009, 2025.

\bibitem{Gao:2013pua}
Long Gao et~al.
\newblock {Attenuation length measurements of a liquid scintillator with
  LabVIEW and reliability evaluation of the device}.
\newblock {\em Chin. Phys. C}, 37:076001, 2013.

\bibitem{Yin:2020elg}
Xiang-Wei Yin et~al.
\newblock {Precise measurement of attenuation length of the JUNO liquid
  scintillator}.
\newblock {\em Radiat. Detect. Technol. Methods}, 4(3):312--318, 2020.

\bibitem{FPGA:1588}
Davide Pedretti et~al.
\newblock Nanoseconds timing system based on ieee 1588 fpga implementation.
\newblock {\em IEEE Transactions on Nuclear Science}, 66(7):1151, 2019.

\bibitem{Aloisio:TTC}
A.~Aloisio et~al.
\newblock {A FPGA-based emulation of the Timing Trigger and control receiver
  for the LHC experiments}.
\newblock In {\em {2011 IEEE Nuclear Science Symposium and Medical Imaging
  Conference}}, page 794, 2011.

\bibitem{Aloisio:RMU}
A.~Aloisio et~al.
\newblock {The Reorganize and Multiplex (RMU) front-end trigger optical bridge
  for the JUNO experiment}.
\newblock In {\em {2017 IEEE Nuclear Science Symposium and Medical Imaging
  Conference}}, 10 2017.

\bibitem{JUNO:2025dfn}
Jilei Xu et~al.
\newblock {Instrumentation of JUNO 3-inch PMTs}.
\newblock 10 2025.

\bibitem{WALKER2026171022}
Pablo~Walker et~al.
\newblock The high voltage splitter board for the juno spmt system.
\newblock {\em Nucl. Instrum. Meth. A}, 1082:171022, 2026.

\bibitem{JUNO:2020orn}
Selma Conforti et~al.
\newblock {CATIROC: an integrated chip for neutrino experiments using
  photomultiplier tubes}.
\newblock {\em JINST}, 16(05):P05010, 2021.

\bibitem{JUNO:DAQ}
Chao Chen et~al.
\newblock {Design and Development of JUNO DAQ Data Flow Software}.
\newblock {\em IEEE Trans. Nucl. Sci.}, 72(3):339--347, 2025.

\bibitem{Zou:2015ioy}
J.~H. Zou et~al.
\newblock {SNiPER: an offline software framework for non-collider physics
  experiments}.
\newblock {\em J. Phys. Conf. Ser.}, 664(7):072053, 2015.

\bibitem{Li:2018fny}
Kaijie Li et~al.
\newblock {GDML based geometry management system for offline software in JUNO}.
\newblock {\em Nucl. Instrum. Meth. A}, 908:43--48, 2018.

\bibitem{JUNO:EDM}
Teng Li et~al.
\newblock {Design and Development of JUNO Event Data Model}.
\newblock {\em Chin. Phys. C}, 41(6):066201, 2017.

\bibitem{brun1997root}
Rene Brun and Fons Rademakers.
\newblock Root—an object oriented data analysis framework.
\newblock {\em Nucl. Instrum. Meth. A}, 389(1-2):81--86, 1997.

\bibitem{You:2017zfr}
Z.~You, K.~Li, Y.~Zhang, J.~Zhu, T.~Lin, and W.~Li.
\newblock {A ROOT Based Event Display Software for JUNO}.
\newblock {\em JINST}, 13(02):T02002, 2018.

\bibitem{Zhu:2018mzu}
Jiang Zhu, Zhengyun You, Yumei Zhang, Ziyuan Li, Shu Zhang, Tao Lin, and
  Weidong Li.
\newblock {A method of detector and event visualization with Unity in JUNO}.
\newblock {\em JINST}, 14(01):T01007, 2019.

\bibitem{Song:2025pnt}
Tian-Zi Song, Kai-Xuan Huang, Yu-Jie Zeng, Ming-Hua Liao, Xue-Sen Wang, Yu-Mei
  Zhang, and Zheng-Yun You.
\newblock {Detector description conversion and visualization in Unity for high
  energy physics experiments}.
\newblock {\em Front. Phys. (Beijing)}, 21(2):26201, 2026.

\bibitem{Tsaregorodtsev:2008zz}
A.~Tsaregorodtsev et~al.
\newblock {DIRAC: A community grid solution}.
\newblock {\em J. Phys. Conf. Ser.}, 119:062048, 2008.

\bibitem{calib:laser}
Yuanyuan Zhang et~al.
\newblock {Laser Calibration System in JUNO}.
\newblock {\em JINST}, 14(01):P01009, 2019.

\bibitem{Takenaka:2024ctk}
Akira Takenaka et~al.
\newblock {Customized calibration sources in the JUNO experiment}.
\newblock {\em JINST}, 19(12):P12019, 2024.

\bibitem{hkk::flasher:private-communication}
Japan Hamamatsu Photonics~K.K.
\newblock Private communication, 2025.

\bibitem{nnvt::flasher:private-communication}
China Northern Night Vision Technology~Co.
\newblock Private communication, 2025.

\bibitem{Kalousis:2019lua}
L.~N. Kalousis, J.~P. A.~M. de~Andr{\'e}, E.~Baussan, and M.~Dracos.
\newblock {A fast numerical method for photomultiplier tube calibration}.
\newblock {\em JINST}, 15(03):P03023, 2020.

\bibitem{PiLas}
{Advanced Laser Diode System A.L.S. GmbH}.
\newblock {\em PiLas - Picosecond Diode Laser}.
\newblock Berlin, Germany, \quad.

\bibitem{JUNO:energy-reso:2025}
Angel Abusleme et~al.
\newblock {Prediction of Energy Resolution in the JUNO Experiment}.
\newblock {\em Chin. Phys. C}, 49(1):013003, 2025.

\bibitem{Wu:2018zwk}
Wenjie Wu, Miao He, Xiang Zhou, and Haoxue Qiao.
\newblock {A new method of energy reconstruction for large spherical liquid
  scintillator detectors}.
\newblock {\em JINST}, 14(03):P03009, 2019.

\bibitem{Liu:2018fpq}
Qin Liu, Miao He, Xuefeng Ding, Weidong Li, and Haiping Peng.
\newblock {A vertex reconstruction algorithm in the central detector of JUNO}.
\newblock {\em JINST}, 13(09):T09005, 2018.

\bibitem{Li:2021oos}
Ziyuan Li et~al.
\newblock {Event vertex and time reconstruction in large-volume liquid
  scintillator detectors}.
\newblock {\em Nucl. Sci. Tech.}, 32(5):49, 2021.

\bibitem{Huang:2021baf}
Guihong Huang et~al.
\newblock {Improving the energy uniformity for large liquid scintillator
  detectors}.
\newblock {\em Nucl. Instrum. Meth. A}, 1001:165287, 2021.

\bibitem{Huang:2022zum}
Gui-hong Huang, Wei Jiang, Liang-jian Wen, Yi-fang Wang, and Wu-Ming Luo.
\newblock {Data-driven simultaneous vertex and energy reconstruction for large
  liquid scintillator detectors}.
\newblock {\em Nucl. Sci. Tech.}, 34(6):83, 2023.

\bibitem{Takenaka:2025hgi}
Akira Takenaka et~al.
\newblock {Neutron source-based event reconstruction algorithm in large liquid
  scintillator detectors}.
\newblock {\em Eur. Phys. J. C}, 85(10):1097, 2025.

\bibitem{Huang:2017abb}
Yongbo Huang et~al.
\newblock {The Flash ADC system and PMT waveform reconstruction for the Daya
  Bay Experiment}.
\newblock {\em Nucl. Instrum. Meth. A}, 895:48--55, 2018.

\bibitem{DayaBay:2019fje}
D.~Adey et~al.
\newblock {A high precision calibration of the nonlinear energy response at
  Daya Bay}.
\newblock {\em Nucl. Instrum. Meth. A}, 940:230--242, 2019.

\bibitem{Borexino:2013zhu}
G.~Bellini et~al.
\newblock {Final results of Borexino Phase-I on low energy solar neutrino
  spectroscopy}.
\newblock {\em Phys. Rev. D}, 89(11):112007, 2014.

\bibitem{DoubleChooz:2019qbj}
H.~de~Kerret et~al.
\newblock {Double Chooz $\theta_{13}$ measurement via total neutron capture
  detection}.
\newblock {\em Nature Phys.}, 16(5):558--564, 2020.

\bibitem{berger1999ESTAR}
J.~Berger Martin\, J.\~S. Coursey, M.\~A. Zucker, and J.~Chang.
\newblock Estar, pstar, and astar: Computer programs for calculating
  stopping-power and range tables for electrons, protons, and helium ions
  (version 1.21).
\newblock \url{http://physics.nist.gov/Star}, 1999.
\newblock Last updated July 2017; Accessed 1 Nov 2025.

\bibitem{Wang:2023iul}
Li~Wang, Jilei Xu, Shuxiang Lu, Haoqi Lu, Zhimin Wang, Min Li, Sibo Wang,
  Changgen Yang, Yongpeng Zhang, and Yichen Zheng.
\newblock {A novel design for 100 meter-scale water attenuation length
  measurement and monitoring}.
\newblock {\em JINST}, 19(05):P05051, 2024.

\bibitem{Yang:2023rbg}
Zekun Yang et~al.
\newblock {First attempt of directionality reconstruction for atmospheric
  neutrinos in a large homogeneous liquid scintillator detector}.
\newblock {\em Phys. Rev. D}, 109(5):052005, 2024.

\bibitem{Liu:2025fry}
Jiaxi Liu et~al.
\newblock {Neutrino type identification for atmospheric neutrinos in a large
  homogeneous liquid scintillation detector}.
\newblock {\em Phys. Rev. D}, 112(1):012018, 2025.

\bibitem{Perraudin:2018rbt}
Nathana{\"e}l Perraudin, Micha{\"e}l Defferrard, Tomasz Kacprzak, and Raphael
  Sgier.
\newblock {DeepSphere: Efficient spherical Convolutional Neural Network with
  HEALPix sampling for cosmological applications}.
\newblock {\em Astron. Comput.}, 27:130--146, 2019.

\bibitem{dai2021coatnet}
Zihang Dai, Hanxiao Liu, Quoc~V. Le, and Mingxing Tan.
\newblock Coatnet: Marrying convolution and attention for all data sizes, 2021.

\bibitem{HEALPix}
Krzysztof~M. G{\'o}rski et~al.
\newblock {HEALPix}: A framework for high-resolution discretization and fast
  analysis of data distributed on the sphere.
\newblock {\em The Astrophysical Journal}, 622(2):759--771, 2005.

\bibitem{Gaisser:2016uoy}
Thomas~K. Gaisser, Ralph Engel, and Elisa Resconi.
\newblock {\em {Cosmic Rays and Particle Physics}: {2nd Edition}}.
\newblock Cambridge University Press, 6 2016.

\bibitem{Tang:2005zz}
Alfred Tang, Chu Ming-Chung, and Kam-Biu Luk.
\newblock {Muon transport simulation for the neutrino experiment at Daya Bay}.
\newblock 6 2005.

\bibitem{MUSIC1}
P.~Antonioli, C.~Ghetti, E.V. Korolkova, V.A. Kudryavtsev, and G.~Sartorelli.
\newblock A three-dimensional code for muon propagation through the rock:
  Music.
\newblock {\em Astroparticle Physics}, 7(4):357–368, October 1997.

\bibitem{MUSIC2}
V.A. Kudryavtsev.
\newblock Muon simulation codes music and musun for underground physics.
\newblock {\em Computer Physics Communications}, 180(3):339–346, March 2009.

\end{thebibliography}
